\documentclass[traditabstract]{aa}
\usepackage{txfonts}
\usepackage{natbib}
\usepackage{graphicx}

\begin{document}

\title{The NIR \ion{Ca}{II} triplet at low metallicity} 
\subtitle{Searching for extremely low-metallicity stars in classical dwarf galaxies\thanks{Using observations collected at the European Organisation for
Astronomical Research in the Southern Hemisphere, Chile proposal 171.B-0588.}}

\author{Else Starkenburg \inst{1} 
\and Vanessa Hill \inst{2} 
\and Eline Tolstoy \inst{1}
\and Jonay I. Gonz\'alez Hern\'andez \inst{3,4,5} 
\and Mike Irwin \inst{6} 
\and Amina Helmi \inst{1} 
\and Giuseppina Battaglia \inst{7} 
\and Pascale Jablonka \inst{4,8} 
\and Martin Tafelmeyer \inst{8} 
\and Matthew Shetrone \inst{9} 
\and Kim Venn \inst{10}
\and Thomas de Boer \inst{1}}

\institute{Kapteyn Astronomical Institute, University of Groningen, P.O. Box 800, 9700 AV Groningen, the Netherlands, email: else@astro.rug.nl 
\and Laboratoire Cassiop\'ee UMR 6202, Universit\'e de Nice Sophia-Antipolis, CNRS, Observatoire de la C\^ote d'Azur, France
\and CIFIST Marie Curie Excellence Team
\and GEPI, Observatoire de Paris, CNRS, Universit\'e Paris Diderot ; Place Jules Janssen, 92190 Meudon, France
\and Dpto. de Astrof{\'\i}sica y Ciencias de la Atm\'osfera, Facultad de Ciencias F{\'\i}sicas, Universidad Complutense de Madrid, E-28040 Madrid, Spain
\and Institute of Astronomy, University of Cambridge, Madingley Road, Cambridge CB03 0HA, UK
\and European Organization for Astronomical Research in the Southern Hemisphere; Karl-Schwarzschild-Strasse 2, 85748 Garching, Germany
\and Observatoire de Gen\`eve, Laboratoire d'Astrophysique de l'Ecole Polytechnique F\'ed\'erale de Lausanne (EPFL), CH-1290 Sauverny, Switzerland
\and University of Texas, McDonald Observatory, HC75 Box 1337-McD, Fort Davis, TX 79734, USA 
\and Department of Physics and Astronomy, University of Victoria, 3800 Finnerty Road, Victoria, BC, V8P 1A1, Canada }

\date{Received 27 November 2009 / Accepted 15 January 2010}

\abstract{The NIR \ion{Ca}{II} triplet absorption lines have proven to be an important tool for
quantitative spectroscopy of individual red giant branch stars in the Local Group, providing a better
understanding of metallicities of stars in the Milky Way and dwarf galaxies and
thereby an opportunity to constrain their chemical evolution
processes. An interesting puzzle in this field is the significant lack of extremely metal-poor
stars, below [Fe/H]=--3, found in classical dwarf galaxies around the
Milky Way using this technique. The question arises whether
these stars are really absent, or if the empirical \ion{Ca}{II} triplet method
used to study these systems is biased in the low-metallicity
regime. Here we present results of synthetic spectral analysis of the
\ion{Ca}{II} triplet, that is focused on a better understanding of
spectroscopic measurements of low-metallicity giant stars. Our results start to deviate strongly from the widely-used and linear empirical calibrations at [Fe/H]$<$--2. We provide
a new calibration for \ion{Ca}{II} triplet studies which is valid for --0.5$\geq$[Fe/H]$\geq$--4. We subsequently apply this new calibration to current
data sets and suggest that the classical dwarf galaxies are not so
devoid of extremely low-metallicity stars as was previously thought.}


\maketitle

\section{Introduction and Outline}

To understand galaxy evolution it is critical to understand how the metallicities of stars in physically different environments develop with time. Low- and high-resolution spectroscopic studies of individual stars in the Milky Way and dwarf galaxies are crucial in this context since they can provide measurements of overall metallicity and detailed abundances of a range of elements \citep[e.g.,][and references therein]{free02}. Studies of this kind have revealed interesting differences between the stars in the Milky Way halo and the dwarf spheroidal galaxies (dSphs) surrounding it \citep[][and references therein]{tols09}. An intriguing population of stars to study in different environments are the extremely metal-poor stars ([Fe/H]$\leq-3$). The existence, or lack of, these stars reveals valuable information about the chemical evolution history of a galaxy, as they represent the most pristine (and probably thus the oldest) stars in the system. For example, in a scenario where there are no low-metallicity stars, theories invoking some kind of pre-enrichment prior to the star formation epoch become more plausible. In general, detailed chemical abundances of old stars can provide valuable information about chemical evolution processes as they are likely to be polluted by only a single or very few supernova. A comparison of low-metallicity tails in different galaxies also provides information about the origin and evolution of systems with different masses and different star formation histories.

From relatively time-consuming but insightful high-resolution (HR) studies of small samples of bright red giant branch (RGB) stars in nearby dSphs we know that the chemical signatures of individual elements in the dSph stars can be distinct from the stars in the Galaxy \citep[e.g.,][]{shet01,shet03,tols03,venn04,fulb04,koch08a,koch08b,cohe09,aoki09,freb10}. So far all high-resolution observations of extremely metal-poor stars in dSphs have been studied in high-resolution as part of follow-up studies on previous low-resolution samples \citep[][Tafelmeyer et al. in prep.]{cohe09,aoki09,freb09,freb10}. These stars are still sparse and at the moment mostly found in the ultra-faint dwarf galaxies. The high-resolution studies of classical dSphs which are not follow-up studies, have first targeted the inner galactic regions, which are often more metal-rich, and are hence not optimized to specifically target metal-poor stars.

Additionally, recent low-resolution (LR) studies enable us to determine overall metallicity estimates for much larger samples of RGB stars in both classical dSphs \citep[e.g.,][]{sunt93,tols04,pont04,batt06,muno06,koch07a,koch07b,batt08a,shet09,walk09a,kirb09}, ultra-faint galaxies \citep[e.g.,][]{simo07,kirb08,norr08,walk09b,koch09} and even the more distant and isolated dwarf irregular galaxies \citep[e.g.,][]{leam09}. From the larger numbers of stars studied in the low-resolution studies for the classical dSphs (typically several hundred per galaxy) one would statistically expect to find some RGB stars with [Fe/H]$\leq-3$, if the distribution of metallicities in these systems follows that of the Galactic halo \citep{helm06}. However, one of the compelling results from studies of large samples of RGB stars is a significant lack of stars with metallicities [Fe/H]$\leq-3$ in the classical dSph galaxies  Sculptor, Fornax, Carina and Sextans compared to the metallicity distribution function of the Galactic halo \citep{helm06}. These metallicities are inferred from the line strengths of the \ion{Ca}{II} NIR triplet (CaT) lines. Recently, \citet{kirb09} reported to have found a RGB star in Sculptor with a [Fe/H] value as low as --3.8 using a comparison between spectra and an extensive spectral library. Several extremely low-metallicity stars have already been discovered in the ultra-faint dwarf galaxies using either this technique or other indicators \citep{kirb08,norr08}. In this paper we investigate whether the lack of low-metallicity stars in the classical dwarf galaxies could be a bias due to the \ion{Ca}{II} NIR triplet (CaT) indicator used to determine metallicities from low-resolution spectra.

The CaT lines have been used in studies over a wide range of atmospheric parameters and have been applied to both individual stars and integrated stellar populations in different environments \citep[see][and references therein]{cena01,cena02}. In this paper we focus on the use of the CaT lines to determine metallicities of individual RGB stars. The CaT region of the spectrum has proven to be a powerful tool for metallicity estimates of individual stars. The three CaT absorption lines ($\lambda$8498, $\lambda$8542, and $\lambda$8662 $\AA$), which can be used to determine radial velocities and to trace metallicity (usually taken as [Fe/H]), are so broad that they can be measured with sufficient accuracy at a moderate resolution. As was already noted in pioneering work \citep[e.g.,][]{arma88,olsz91,arma91}, there are numerous additional advantages to the use of the CaT lines as metallicity indicator. For instance, the calcium abundances are expected to be largely representative of the primordial abundances of the star, since, contrary to many other elements, they are thought to be unaltered by nucleosynthesis processes in intermediate- and low-mass stars on the RGB \citep{ivan01,cole04}. Also, the NIR wavelength region is convenient: in that the red giants emit more flux in this part of the spectrum than in the blue and the spectrum is very flat, which facilitates the definition of the continuum level to measure the equivalent widths of the line. 

However, the breadth of the lines also has disadvantages. Because the lines are highly saturated their strength, especially in the core of the line, depends strongly on the temperature structure of the upper layers of the photosphere and chromosphere of the star, which means that complicated non-local thermodynamic equilibrium (non-LTE) physics has to be used to model the line correctly \citep{cole04}. Also, the lines do not provide a direct measurement of [Fe/H], although it has been shown that [Fe/H] affects the equivalent width of the lines more than Ca \citep{batt08b}. The abundance of Ca and other elements do still affect the equivalent widths which will not always trace only [Fe/H] \citep{rutl97b}. 

Early investigations of the CaT concentrated mainly on their sensitivity to surface gravity \citep{spin69,spin71,cohe78,jone84}. It was realized by \citet{arma88} that a more metal-rich RGB star should have stronger CaT lines, because it has both a greater abundance of Fe (and also Ca) in its atmosphere and a lower surface gravity. They empirically proved this relation by measuring for the CaT lines their integrated equivalent width (EW) in several globular clusters with known [Fe/H]. Applying this method to individual RGB stars, \citet{olsz91} noticed that the metallicity sensitivity of the CaT line index is improved by plotting it as a function of the absolute magnitude of the star. At a fixed absolute magnitude, higher metallicity RGB stars will have lower gravity and lower temperatures, which both strengthen the CaT lines. A further development of the method \citep{arma91} was to plot the equivalent width as a function of the height of the RGB stars above the horizontal branch (HB) in V-magnitude ($V-V_{\textnormal{\scriptsize{HB}}}$). In this way, the requirements of a distance scale and a well-determined reddening are avoided. They found a linear relation between [Fe/H] and a ``reduced equivalent width'' (W$^{'}$), which incorporates both the equivalent width of the two strongest lines at $\lambda8542$ and $\lambda8662\ \AA$ (also written as EW$_{2}$ and  EW$_{3}$ in the remaining of this paper) and $V-V_{\textnormal{\scriptsize{HB}}}$. This enables a direct comparison between RGB stars of different luminosities. This method has been extensively tested and proven to work on large samples of individual RGB stars in globular clusters \citep[e.g.,][]{rutl97a,rutl97b}. 

Additionally, \citet{cole04} showed that the effect of different ages of RGB stars is a negligible source of error for metallicities derived from the CaT index. This paved the way for the use of the CaT metallicity indicator on populations of RGB stars which are not coeval, among which are the Local Group dwarf galaxies. A direct detailed comparison between the low-resolution CaT metallicities and high-resolution measurements for large samples of RGB stars in the nearby dwarf galaxies Fornax and Sculptor is given by \citet{batt08b}. They concluded that the CaT - [Fe/H] relation (calibrated on globular clusters) can be applied with confidence to RGB stars in composite stellar populations over the range $-2.5 < $[Fe/H]$ < -0.5$. 

In this paper we study the behavior of the CaT for [Fe/H] $< -2.5$, comparing stellar atmosphere models with observations. We have determined a new calibration including this low-metallicity regime. The paper is organized as follows. In Sect. \ref{CaTlowmet} we further clarify and physically motivate the need for a new calibration at lower metallicities using simple synthetic spectra modeling. In Sect. \ref{models} we describe our grid of synthetic spectra we use to investigate the CaT lines. We then analyze and calibrate this grid in two regimes: --2.0 $\geq$ [Fe/H] $\geq$ --0.5 in Sect. \ref{highmet}, and [Fe/H] $\leq$ --2.5 in Sect. \ref{lowmet}. In Sect. \ref{newcalib} we present a new calibration valid for both metallicity regimes. Additionally, we investigate the role of $\alpha$ elements in Sect. \ref{alpha} and the implications of the new calibration on the results of the Dwarf Abundances and Radial velocities Team (DART) survey \citep{tols06} in Sect. \ref{implications}. 

\section{The CaT at low metallicity}\label{CaTlowmet}

The so-called CaT empirical relation connects a linear combination of the equivalent widths of the CaT lines (the exact form can vary between different authors) and the absolute luminosity of the star, often expressed in terms of the height of the star above the HB of the system, to its [Fe/H] value. The empirical relation described in \citet{batt08b} in their equations 16 and 11 are given as  equations \ref{emprel1} and \ref{emprel2} below.

\begin{equation}
[\textnormal{Fe}/\textnormal{H}] = -2.81(\pm0.16)+0.44(\pm0.04)W^{'}  
\label{emprel1}
\end{equation}
where: 
\begin{equation}
W^{'} = EW_{2} + EW_{3} + 0.64(\pm0.02)(V-V_{\textnormal{\scriptsize{HB}}})
\label{emprel2}
\end{equation}

At present this linear relation between metallicity and CaT equivalent widths is also used to infer metallicities for stars outside the calibrated regime ($-2.5 < $[Fe/H]$ < -0.5$). However, the assumption that the relation continues outside of this regime has not been accurately checked. It is very clear that the linear relation \textit{cannot} hold down to extremely low metallicities, since at a certain point Eqs.\ref{emprel1} and \ref{emprel2} infer that the equivalent widths will have negative values. Negative equivalent widths for absorption lines is obviously physically meaningless. 

\begin{figure}
\resizebox{\hsize}{!}{\includegraphics{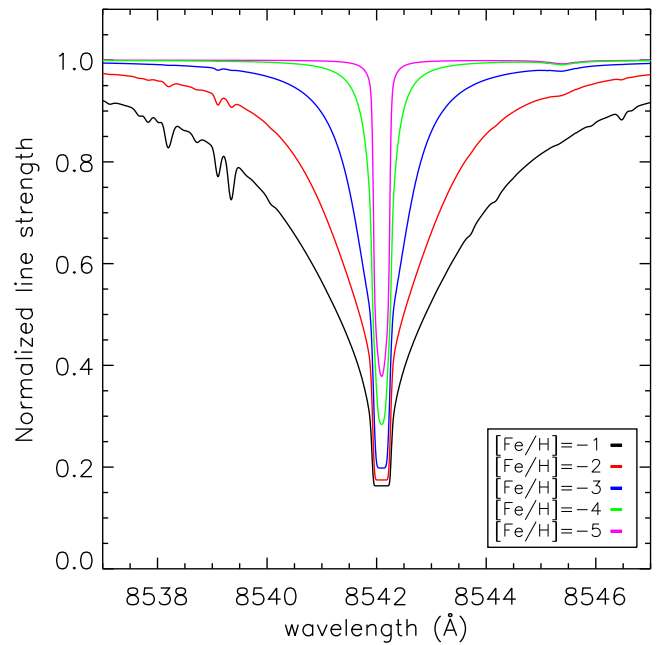}}
\caption{The change of the shape of the CaT line at $\lambda$ 8542 \AA with changing metallicity (metallicity is decreasing as the line narrows). Non-LTE effects are not taken into account. These lines are obtained from synthetic spectra from MARCS models with T$_{\textnormal{\scriptsize{eff}}}$ = 4500 K and log(g) = 1.5 as described in the text. The flattening in the core at higher metallicities arises because the models do not cover small enough optical depths in the outer layers, where the core of the line is formed. This is a problem within the modeling itself, and not related to the non-LTE effects which are described in Sect. \ref{non-LTE}. Because of the broad wings, the area of the core of the line that is missing is a negligible fraction of the total equivalent width. \label{justCaT_models}}
\end{figure}

It is also expected that the shape of the lines, and therefore the relation of equivalent width with metallicity, changes as the metallicity drops. The equivalent width of the CaT lines is dominated at higher metallicities by the wings of the line. At lower metallicities these wings weaken and/or even disappear and the equivalent width of the line becomes more dominated by its core, as illustrated in Fig. \ref{justCaT_models} from synthetic spectra. Since the core and the wings of the lines originate from different parts of the curve of growth the behavior of both is unlikely to be described by one linear equation. The change in relative strength of the core and wings of the line changes the sensitivity of the line to the Ca abundance and to the overall metallicity in general. This understanding of the physical process motivates our re-calibration of the relation between CaT equivalent width and [Fe/H] at lower metallicities.

\section{A grid of Models}\label{models}
In order to investigate and quantify the behavior of the CaT lines at low metallicity we define a grid of model spectra created using the publicly available and recently revised version of the (OS)MARCS models \citep[e.g.,][]{gust08,plez08} and the Turbospectrum program \citep{alva98} updated consistently with the latest release of MARCS \citet{gust08}. The model spectra cover a range of effective temperatures, gravities, metallicities ([Fe/H]) and enhancements of the $\alpha$ elements (taken as O, Ne, Mg, Si, S, Ar, Ca, and Ti in the MARCS models \citep{gust08}). The models use a 1D spherical symmetric approach. Our linelist is created using the Vienna Atomic Line Database (VALD) for the atomic species \citep[e.g.,][]{kupk00}. Additionally we also model the contribution from CN molecular lines \citep[][Plez priv comm.]{spit05} and  TiO moleculer lines in the cooler (T$_{\textnormal{\scriptsize{eff}}} < 4000 K$) models \citep{plez98}.

\begin{table*}
\caption{The parameters for the grid of models used.}
\label{modelgrid}
\centering
\begin{tabular}{lrlr}
\hline\hline
\multicolumn{1}{c}{[Fe/H]} & \multicolumn{1}{c}{[$\alpha$/Fe]} & \multicolumn{1}{c}{T$_{\textnormal{\scriptsize{eff}}}$} & \multicolumn{1}{c}{log(g)}\\
\hline
\multicolumn{4}{c}{MARCS model atmospheres}\\
\hline
--0.25 & +0.0, +0.1 & 3800, 3900, 4000, 4250, 4500, 4750, 5000 & 0.5, 1.0, 1.5, 2.0, 2.5 \\
--0.50 & +0.0, +0.2 & 3800, 3900, 4000, 4250, 4500, 4750, 5000 & 0.5, 1.0, 1.5, 2.0, 2.5 \\
--0.75 & +0.0, +0.3 & 3800, 3900, 4000, 4250, 4500, 4750, 5000 & 0.5, 1.0, 1.5, 2.0, 2.5 \\
--1.00, --1.50, --2.00 & +0.0, +0.4 & 3800, 3900, 4000, 4250, 4500, 4750, 5000 & 0.5, 1.0, 1.5, 2.0, 2.5 \\
--2.50, --3.00 & +0.4 & 3800, 3900, 4000, 4250, 4500, 4750, 5000 & 0.5, 1.0, 1.5, 2.0, 2.5 \\
--4.00, --5.00 & +0.4 & 3800, 3900, 4000, 4250, 4500, 4750, 5000 & 1.0, 1.5, 2.0, 2.5 \\
\hline
\multicolumn{4}{c}{Interpolated model atmospheres}\\
\hline
--1.25, --1.75 & & 4125, 4375, 4625, 4875  & 0.75, 1.25, 1.75, 2.25 \\
--2.25, --2.75  & & 4125, 4375, 4625, 4875 & 0.75, 1.25, 1.75, 2.25 \\
--3.50, --4.50 & & 4125, 4375, 4625, 4875 &        1.25, 1.75, 2.25 \\
\end{tabular}
\end{table*}

\subsection{Parameters}
The parameters [Fe/H], [$\alpha$/Fe], T$_{\textnormal{\scriptsize{eff}}}$ and log(g) for MARCS model atmospheres in our grid are given in Table \ref{modelgrid}. For the upper part of the table all the models are provided on the web by the MARCS team. We interpolated to a finer grid, including the values given in the lower part of Table \ref{modelgrid}, using the interpolation tool provided on the MARCS website by T. Masseron. All atmospheric models have a mass of $1M_{\odot}$ and microturbulent velocities ($\xi$) of 2 km s$^{-1}$. In calculating the synthetic spectra with Turbospectrum we use microturbulent velocities which vary slightly dependent on the gravity of the atmospheric model in accordance with observed variations of microturbulent velocities in halo stars. The values we use range from $\xi=2.3$ km s$^{-1}$ for the tip of the RGB (log(g) = 0.5) to $\xi=1.7$ km s$^{-1}$ near the HB (log(g)=2.5) in accordance with the results from \citet{bark05}. This slight change in microturbulence can significantly alter the equivalent width of the two strongest CaT lines in the metal-poor regime and the difference will be largest at the tip of the RGB.

\subsection{Non-LTE effects}\label{non-LTE}

Although all model atmospheres and synthetic spectra are calculated assuming local thermodynamic equilibrium (LTE), we take effects of departures from LTE into account, because they can be significant since the lines are so highly saturated. The effect of departures from LTE on these lines was first investigated by \citet{jorg92}, but only for [Fe/H] $\geq -1$. A more extensive study is performed by \citet{mash07}. They note that in the CaT the non-LTE effects are revealed only in the Doppler core, which is strengthened.  

\begin{figure*}
\sidecaption
\includegraphics[width=12cm]{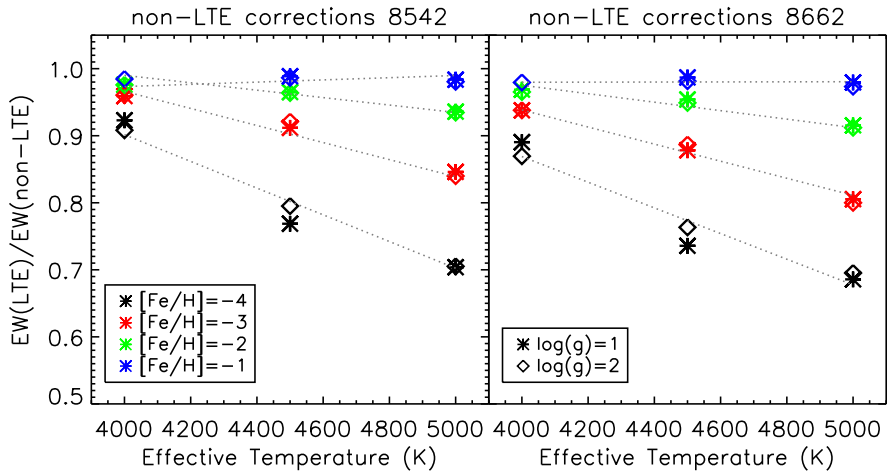}
\caption{The ratio of LTE to full non-LTE equivalent widths for the different models for the second (8542\AA) CaT line in the left panel and the third (8662\AA) CaT line in the right panel. For the models with [Fe/H]$<$ --2, non-LTE effects make significant contributions to the line-strength.  The gray dotted lines indicate a simple fit to these values, using the metallicity and temperature of the model as input parameters.\label{nlte}}
\end{figure*}

Because the broad wings of the CaT lines are decreasing significantly with metallicity, as shown in Fig. \ref{justCaT_models}, the effect of departures from LTE, which only affect the core, have most impact on the equivalent width determination at low metallicities. Therefore, it is crucial to take non-LTE effects into account in order to understand the behavior of the CaT lines at low metallicity. We perform non-LTE calculations using the model atom presented by \citet{mash07}, which contains 63 levels of \ion{Ca}{I}, 37 levels of \ion{Ca}{II} and the ground state of \ion{Ca}{III}. Non-LTE level populations and synthetic spectra were determined with recent versions of the codes DETAIL and SURFACE \citep{gidd81,butl85}. We chose ATLAS9 atmospheric models \citep{kuru93} as the input models in the non-LTE computations. Thus, we computed ATLAS9 model atmospheres exactly for a given set of stellar parameters and metallicities using a Linux version of the ATLAS9 code \citep{sbor04}, and adopting the new Opacity Distribution Functions (ODFs) of \citet{cast03}. We also adopted $S_{\rm H}=0.1$ as the scaling factor of the inelastic collisions with hydrogen atoms in the non-LTE computations. For further details on the non-LTE computations we refer to \citet{mash07}.

 We determined the equivalent widths by integrating normalized fluxes of the broad \ion{Ca}{II} triplet non-LTE profiles at 8542 and 8662\AA. Figure \ref{nlte} shows the ratio of LTE EWs to non-LTE EWs from this modeling for both these CaT lines in separate panels. At high temperature and [Fe/H]=--4.0 the ratio goes down  in both lines to $\sim$0.7, which means that just $\sim$70\% of the line is modeled in LTE and non-LTE effects are thus very important. We determine a best-fitting relation as a function of metallicity and temperature of the model using the IDL function MPFIT2DFUN \citep{mark09}, which performs a Levenberg-Marquardt least-squares fit to a 2-D function, in combination with a statistical F-test to identify the best fit. The best-fitting relations obtained separately for the two strongest CaT lines are shown as dashed gray lines in Fig. \ref{nlte} and given in Eqs. \ref{eqnlte1} and \ref{eqnlte2}.

\begin{eqnarray}
\textnormal{EW(8542\AA)}_{\textnormal{\scriptsize{LTE}}}/\textnormal{EW(8542\AA)}_{\textnormal{\scriptsize{non-LTE}}} = 0.563 + 0.397\frac{\textnormal{T}_{\textnormal{\scriptsize{eff}}}}{4500 (\textnormal{K})} \nonumber \\
- \ (0.365 - 0.323\frac{\textnormal{T}_{\textnormal{\scriptsize{eff}}}}{4500 (\textnormal{K})})[\textnormal{Fe/H}] - 0.0203[\textnormal{Fe/H}]^2 
\label{eqnlte1}
\end{eqnarray}

\begin{eqnarray}
\textnormal{EW(8662\AA)}_{\textnormal{\scriptsize{LTE}}}/\textnormal{EW(8662\AA)}_{\textnormal{\scriptsize{non-LTE}}} = 0.692 + 0.292\frac{\textnormal{T}_{\textnormal{\scriptsize{eff}}}}{4500 (\textnormal{K})} \nonumber \\
- \ (0.301 - 0.288\frac{\textnormal{T}_{\textnormal{\scriptsize{eff}}}}{4500 (\textnormal{K})})[\textnormal{Fe/H}] - 0.0163[\textnormal{Fe/H}]^2 
\label{eqnlte2}
\end{eqnarray}

For our finer grid of synthetic spectra from MARCS model atmospheres described in the previous paragraph we use these equations to determine the ratio of LTE to non-LTE equivalent widths for the two lines in each individual model. By dividing the individual equivalent widths in the grid (which are all calculated in LTE) by the factor for the corresponding line before adding the two strongest lines together, we correct each of the grid points for non-LTE effects.  

\section{The CaT lines at --2.0 $\leq$ [Fe/H] $\leq$ --0.5}\label{highmet}

\subsection{The empirical relation}\label{emprel}
The empirical relation between CaT EW, absolute magnitude and [Fe/H] is very well studied in globular clusters, i.e., in the metallicity range between [Fe/H]$\approx$--2.3 and [Fe/H]$\approx$--0.5. This well-known relation can thus be used to test our synthetic spectra at these metallicities. We make a comparison between the CaT lines from atmospheric models between --0.5 $<$ [Fe/H] $<$ --2.0 and the best fit linear relation for globular clusters as given in Eqs. \ref{emprel1} and \ref{emprel2} in Sect. \ref{CaTlowmet} \citep[from][]{batt08b}.

Some approximations are needed to enable a CaT analysis with synthetic spectra which is comparable to observations. 

1. The synthetic spectra are all degraded to a resolution of R=6500, the resolution of VLT/FLAMES used in Medusa mode with the GIRAFFE LR (LR8) grating as was used in the observational determination of the empirical relations given above. The equivalent widths for the two strongest CaT lines are measured using the same fitting routine as in \citet{batt08b}: a Gaussian fit with a correction which comes from a comparison with a the summed flux contained in a 15$\AA$ wide region centered on each line. The correction is necessary to account for the wings in the strong lines which are distinctly non-Gaussian in shape. Since the CaT lines can have a variety of shapes for a range of metallicities, as shown in Fig. \ref{justCaT_models}, it is in general not advisable to fit them using a single profile. The disadvantage of using just numerical integration of the observed spectra is that there are also some weaker lines in this wavelength range that may vary differently with changing stellar atmospheric parameters than the CaT lines themselves \citep{carr07}. Taking these considerations into account, we thus use a combination of both methods. Weak nearby lines in the spectrum still introduce a small dependence of the measured CaT EW on resolution though, since they are more likely to be absorbed into the Gaussian fit at a lower resolution. This dependence is already present the lowest metallicity in our grid and gets stronger at higher metallicity, due to the increasing prominence of the non-Gaussian wings. 

Some of the more prominent weak lines which can be present in the wings of the two strongest CaT lines are the hydrogen Paschen lines. Their strength mainly depends on the temperature of the star (the hotter the stronger) and also to a lesser extend on its gravity (increasing with decreasing gravities). In the more metal-rich part of our grid the Paschen lines fall within the broad wings of the CaT lines and have a direct effect on the EW measurement of the CaT line. However, within the range of parameters we use in our models, the maximum contribution of the Paschen lines to the CaT EW measured is 39 m\AA, which is negligible compared to the total CaT EW. In the more metal-poor part of the grid the CaT lines are narrower and well separated from the Paschen lines. Nonetheless, the Paschen lines can still influence the CaT EW by effecting the placement of the continuum. Also this effect we find to be negligible at our grid parameters, at maximum the CaT EW is changed by 3.5\%.     

\begin{figure}
\resizebox{\hsize}{!}{\includegraphics{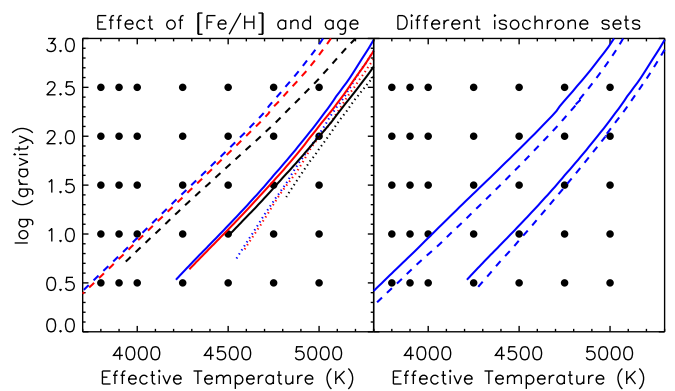}}
\caption{The log(g) and effective temperature space occupied by a selection of the atmospheric models (dots) and theoretical isochrones (lines). In the left panel Yonsei-Yale isochrones are shown for different metallicities ([Fe/H]=-1 dashed lines, [Fe/H]=-2 solid lines, and [Fe/H]=-3 dotted lines) and ages (12 Gyr, 8 Gyr, and 2 Gyr (respectively blue, red and black)). In the right panel the Yonsei-Yale isochrones for 12 Gyr and [Fe/H]=-1 and [Fe/H]=-2 are plotted (solid lines) and overplotted with the same isochrones from BaSTI (dashed lines). All isochrones plotted here assume an alpha-element enhancement of [$\alpha$/Fe]=+0.4. \label{gridandiso}}
\end{figure}  

2. Not all the models in our grid, each of them a particular combination of effective temperature, gravity and metallicity, will represent real stars on the RGB. To determine which models best compare to real stars we use two sets of isochrones, the BaSTI isochrones \citep{piet04} and the Yonsei-Yale set \citep[e.g.,][]{yi01,dema04}. Both sets can be interpolated to obtain exactly the desired metallicity and age for a particular isochrone. The Yonsei-Yale set has the advantage that they go down to [Fe/H] = --3.6 at [$\alpha$/Fe]=+0.4, whereas the lowest value for the BaSTI set is [Fe/H] = --2.6. It is well known that different sets of isochrones do not always give identical results. On top of that, different ages give (slightly) different parameters for the RGB stars. In Fig. \ref{gridandiso} the grid of models in effective temperature versus gravity is plotted along with the relations from the theoretical isochrones. In the two panels the differences due to age, metallicity and the use of different sets of isochrones are illustrated. Based on Fig. \ref{gridandiso} we decide to linearly interpolate in log(g) space between the models that are as close as possible to the isochrones and add another 0.25 in log(g) space on each side to account for uncertainties within the isochrone models as well as the differences between the isochrones of different ages. These uncertainties are shown as error bars on the equivalent widths from synthetic spectra. Note that these represent maximum and minimum values for the equivalent widths and that these errors are systematic. 

3. Because our synthetic grid is not a stellar system, the HB magnitude does not have an obvious meaning. Therefore, to compare our models with empirical relations which require the height above the HB as an input parameter, we have to rely on observational or theoretical relations between M$_{\textnormal{\scriptsize{V}}}$ of the HB (V$_{\textnormal{\scriptsize{HB}}}$) of a system and its metallicity. The M$_{\textnormal{\scriptsize{V}}}$ for each of the grid RGB stars is taken from the isochrones. This value is consistent with the value we get if we calculate M$_{\textnormal{\scriptsize{V}}}$ from M$_{\textnormal{\scriptsize{bol}}}$ using the bolometric correction of giant stars from \citet{alon99}. To determine V$_{\textnormal{\scriptsize{HB}}}$ we use the relation given in \citet{cate08} which is calculated using theoretical models for RR Lyrae stars. Within its uncertainties, this relation is in excellent agreement with observations of the V$_{\textnormal{\scriptsize{HB}}}$ of globular clusters \citep[e.g.,][]{rich05}.

In order to make a fair comparison between the synthetic spectra and the empirical relation, the atmospheric model properties were chosen to be as close as possible to known globular cluster properties. In the atmospheric models we therefore chose to use the alpha-enhanced models with [$\alpha$/Fe]=+0.4, except at the higher metallicities ([Fe/H]$>$--1.0) where the MARCS grid only provides lower [$\alpha$/Fe] to match observations in the Milky Way. Furthermore, for all [$\alpha$/Fe]=+0.4 models, [Ca/Fe] was set to +0.25 in the Turbospectrum program, which even more closely resembles the true values observed in the Galactic halo globular clusters \citep[e.g.,][]{prit05}. For the isochrone set we use an age of 12 Gyr, comparable to measured ages for the globular clusters \citep[e.g.,][]{krau03}. We find that for a range of old ages, between 8 and 15 Gyr, the exact choice of the isochrone age does not significantly affect our results.  

\begin{figure*}
\includegraphics[width=17cm]{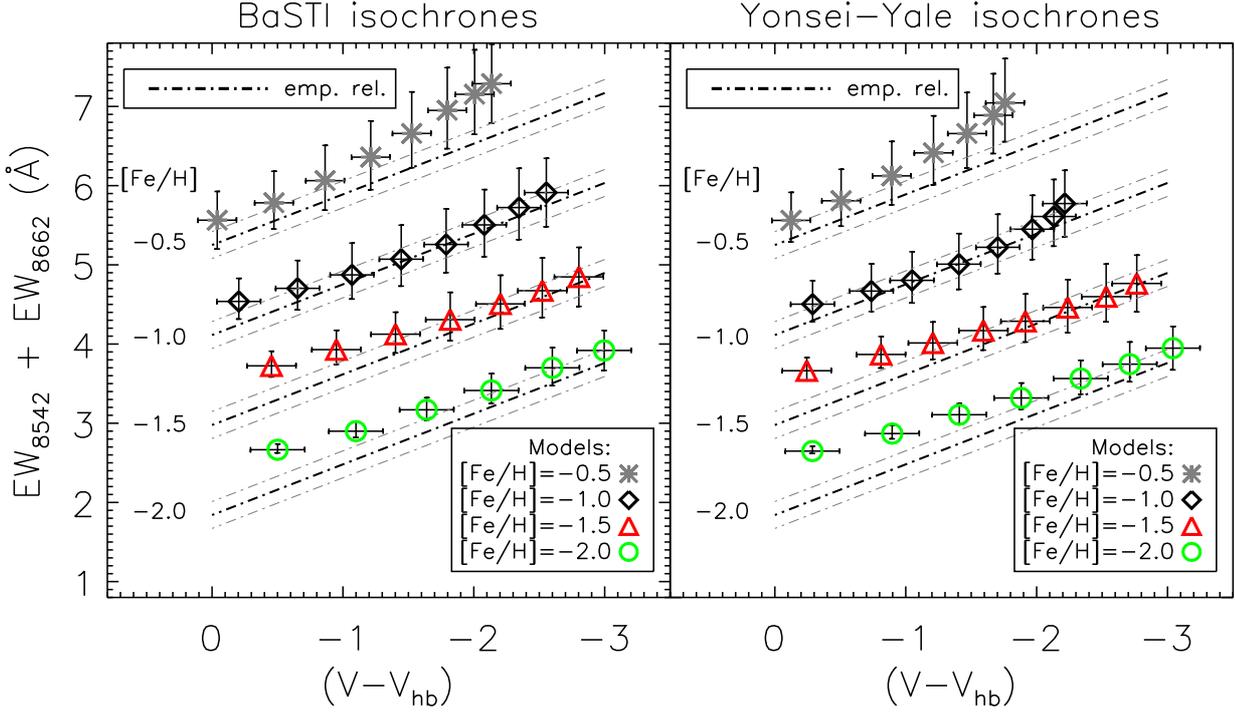}
\caption{We plot equivalent widths of the two strongest CaT lines which are measured in the synthetic spectra (symbols) and the empirical relation from Battaglia et al. (2008b) (black dash-dotted lines) versus (V-V$_{\textnormal{\scriptsize{HB}}}$), calculated as described in the text. For the empirical relation we derived 1$\sigma$ uncertainties from the spread of the individual RGB stars in globular clusters from Battaglia et al. (2008b) around the empirical relation, these are shown as gray dash-dotted lines. The synthetic spectra are calculated for metallicities [Fe/H]=--0.5 (gray asterisks), [Fe/H]=--1.0 (black diamonds), [Fe/H]=--1.5 (red triangles) and [Fe/H]=--2.0 (green circles). The empirical relation is calculated for the same range of metallicities (from top to bottom). The synthetic spectra are matched to the physical properties of RGB stars using the BaSTI isochrones (left panel) and the Yonsei-Yale isochrones (right panel). The error bars denote the uncertainties due to the calculation of HB magnitudes (horizontally) and the log(g)-T$_{\textnormal{\scriptsize{eff}}}$ relation from isochrones (vertically), for which the error bars show a maximum and minimum value. \label{modemp_highmet}}
\end{figure*} 

The successful comparison between our synthetic spectra and the empirical relations, using the BaSTI and the Yonsei-Yale sets of isochrones, is shown in Fig. \ref{modemp_highmet}. The results for the two different sets of isochrones are comparable, although the BaSTI isochrones give a slightly better coverage and agreement at the tip of the RGB. In general, we obtain a good match between the predictions of our synthetic spectra equivalent widths and the empirical relation, especially for the most luminous part of the RGB and at intermediate metallicities where the relation is best studied. At [Fe/H]=--0.5 the match is clearly worse, but also the empirical relation is not well constrained at this metallicity. We also find a larger deviation from the empirical relation closer to the HB. It was already predicted by \citet{pont04} using models that the strength of the CaT lines increases more rapidly for the more luminous part of the RGB. Later, \citet{carr07} reported a nonlinear tendency in the equivalent width versus absolute magnitude relation at fainter magnitudes from an observational study of a large sample of RGB stars in open and globular clusters over a wide range of magnitudes. Recently, \citet{daco09} reported a flattening of the slope of the relation below the HB in two globular clusters. These observations and early predictions appear to confirm the effect we see in our synthetic spectra. This trend is observed to be even stronger below the HB \citep{carr07,daco09}, and clearly shows one has to be very cautious applying any of the [Fe/H]-CaT relations to faint stars, especially below the HB. In this paper, we only focus on the RGB above the HB.

\subsection{Further calibration}
In addition to the comparison with the existing empirical relations, we also calibrate our spectral synthesis models with two (very) well studied examples, namely the Sun using the Kurucz solar flux atlas \citep{kuru84} and Arcturus using the Hinkle Arcturus atlas \citep{hink00}. Although the match for the Sun is very good, the initial comparison between the observational Hinkle Arcturus spectrum and the synthesized spectrum from our models was not satisfactory, especially in the wings of the strongest CaT line ($\lambda$ 8542 \AA). Because this line is extremely broad, it is possible that some of the outer parts of the line were mistakenly taken to be the continuum level during the continuum subtraction. After careful renormalization of the continuum at this wavelength region we were able to get an acceptable match using the abundances for Arcturus from \citet{fulb07}.

\section{The CaT lines at [Fe/H] $<$ --2.5}\label{lowmet}

\subsection{The empirical relation}
\begin{figure}
\resizebox{\hsize}{!}{\includegraphics{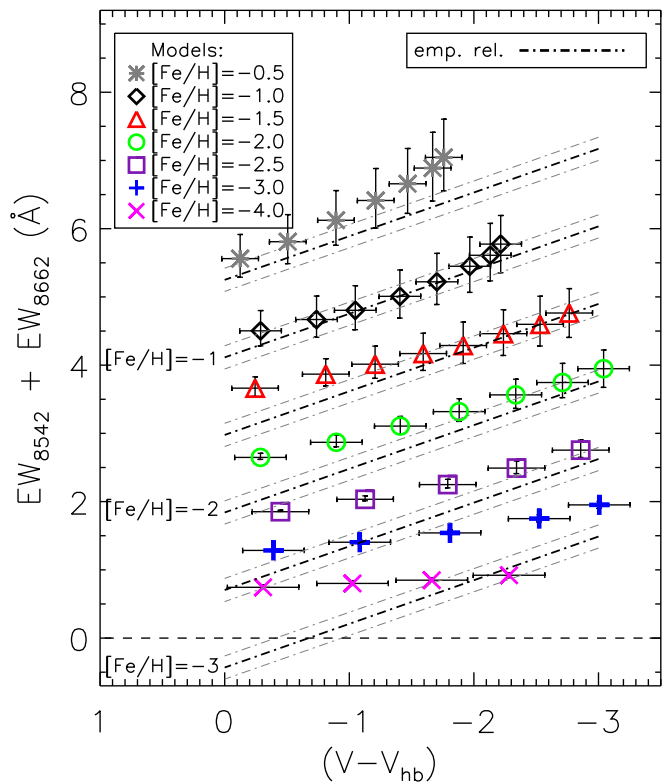}}
\caption{From synthetic analysis we measure the equivalent widths of the two strongest CaT lines as in Fig. \ref{modemp_highmet} extended to lower metallicities. Color coding and symbols for --2.5 $<$ [Fe/H] $<$ --0.5 is the same as in Fig. \ref{modemp_highmet}, the additional metallicities are [Fe/H]=--2.5 (purple squares), [Fe/H]=--3.0 (blue plus signs), and [Fe/H]=--4.0 (pink crosses). The empirical relations (black dash-dotted lines), including 1$\sigma$ uncertainties (gray dash-dotted lines) are shown for [Fe/H]=--0.5, [Fe/H]=--1.0, [Fe/H]=--1.5, [Fe/H]=--2.0, [Fe/H]=--2.5, and [Fe/H]=--3.0 (from top to bottom). \label{modemp_lowmet}}
\end{figure}

Given the success in reproducing the well established calibration of CaT in the range --2.0 $\leq$ [Fe/H] $\leq$ --0.5, we now extend our synthetic spectral analysis down to [Fe/H]=--4.0. The results are shown in Fig. \ref{modemp_lowmet} together with the empirical relation extended linearly to the low-metallicity regime. We use the Yonsei-Yale isochrone set since it extends down to the lowest metallicities\footnote{The Yonsei-Yale isochrone set provides isochrones with [Fe/H]$\geq$--3.6, so we have to use the [Fe/H]=--3.6 isochrone for our [Fe/H]=--4 models. Because at lower metallicities the isochrones come closer together the error introduced by this mismatch is negligible.}. From a comparison of the empirical relation (dash-dotted lines) and the synthetic spectra (colored symbols) in Fig. \ref{modemp_lowmet}, it is obvious that the match with the empirical linear relations breaks down (as expected) at low metallicities, starting from [Fe/H]$<$--2.0. For example, the empirical relation for [Fe/H]=--3.0 lies below the synthetic spectra predictions for [Fe/H]=--4.0 models for fainter RGB stars. 

As can be seen in Fig. \ref{modemp_lowmet}, there is also a trend with the inferred height of a `star' above the HB (e.g., surface gravity and temperature) and the deviation of the `observed' metallicity compared to the metallicity of the model, for the low-metallicity models. This demonstrates the fact that not just the equivalent width, but also the slope of the relation is changing. At lower metallicities, the equivalent width of the line becomes less sensitive to variations in gravity or temperature of the star (i.e., its position on the RGB).  

\begin{figure}
\resizebox{\hsize}{!}{\includegraphics{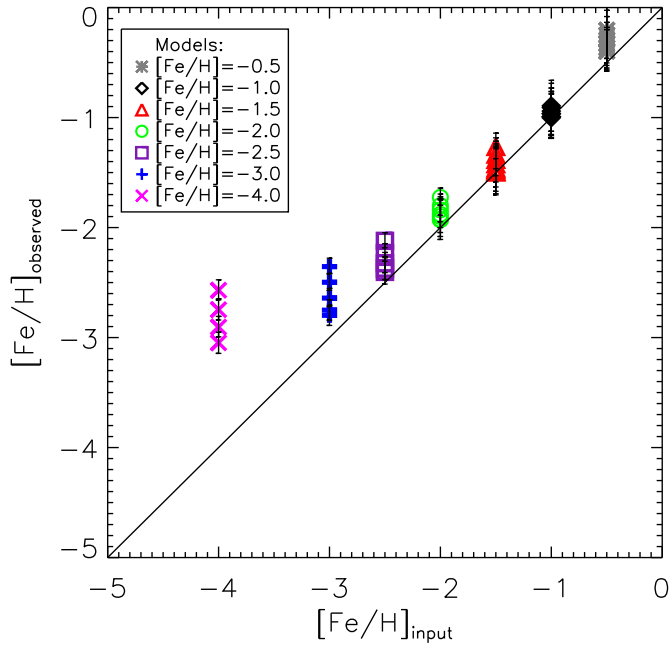}}
\caption{From synthetic analysis, we plot the `real' input value for [Fe/H] in the model versus the 'observed' value of [Fe/H] for the grid of models, obtained by treating the model spectra as if they were observed in the DART program. For each `real' [Fe/H] input value are several points representing RGB stars of the same metallicity at several places along the RGB. Error bars are calculated from the uncertainties shown in Fig. \ref{modemp_highmet} and described in the text. If the CaT method would work perfectly all models should fall (within their uncertainties) on the one-to-one relation shown by the solid black line. Clearly there is an increasing deviation starting at [Fe/H]$<$--2. Same symbols and color coding used as in Fig. \ref{modemp_lowmet}.\label{femodelvsfeCaT}}
\end{figure}

The mismatch between the extended empirical relation and the synthetic modeling predictions at low metallicities is emphasized in Fig. \ref{femodelvsfeCaT} where the input metallicity of the models is plotted versus the metallicity obtained from the empirical relation as given in Eqs. \ref{emprel1} and \ref{emprel2}. From Fig. \ref{femodelvsfeCaT}, we obtain valuable insight into how extremely low-metallicity spectra would appear in, for instance, the DART sample of classical dwarf galaxies. While some of the models with input [Fe/H] $\leq -2.5$ are correctly reproduced by the synthetic spectral method, there are also examples where the metallicity is seriously overestimated. The linear empirical relation thus offers no means to discriminate between very low or extremely low metallicity RGB stars.   

\section{A new calibration}\label{newcalib}

\begin{figure}
\resizebox{\hsize}{!}{\includegraphics{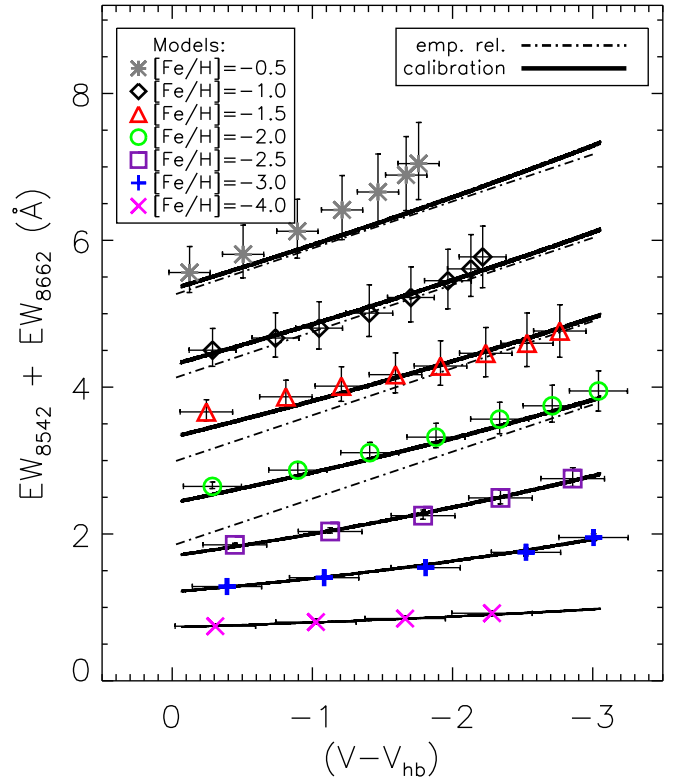}}
\caption{Same as Fig. \ref{modemp_lowmet}, but with our new calibration overplotted as thick solid black lines.\label{newcalibfig_a}}
\end{figure}

\begin{figure}
\resizebox{\hsize}{!}{\includegraphics{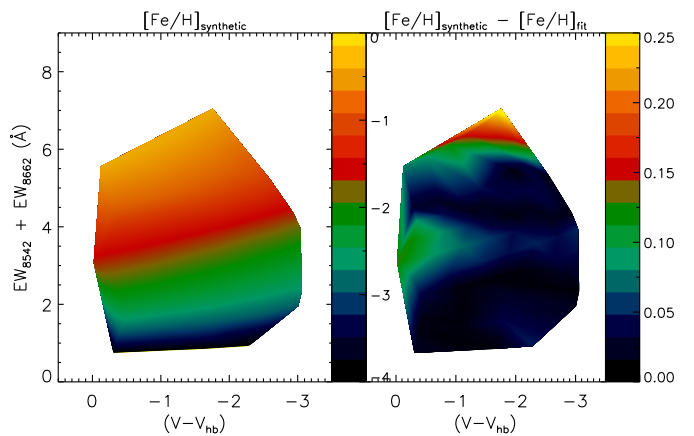}}
\caption{Two density plots showing the [Fe/H] values for the complete set of models (left panel) and the difference between the [Fe/H] values of the models and their calculated [Fe/H] using the new calibration (right panel).\label{newcalibfig_b}}
\end{figure}   

To re-calibrate the relation between the CaT equivalent width and metallicity for the low-metallicity regime, we use the equivalent widths obtained from our synthetic spectra. In the relation obtained by \citet{batt08b}, there are three free parameters to fit the slope of the relation as a function of height above the HB, expressed in W$^{'}$, and a linear relation between W$^{'}$ and [Fe/H]. To extend this calibration we expect to need at least two more parameters to fit the dominant features of the low-metallicity models in one relation: the changing slope with [Fe/H] and the changing offset between lines of equal metallicity. As input into our fitting routine we use the results from the synthetic spectra grid from models described in Table \ref{modelgrid} with [$\alpha$/Fe] = +0.4 (and [Ca/Fe] set to +0.25). We use the Yonsei-Yale set of isochrones, because only those provide the low metallicities needed for the analysis. For the fitting we use the IDL function MPFIT2DFUN \citep{mark09} to fit a plane relating the absolute magnitude and equivalent widths of the synthetic models to their metallicity and an F-test to distinguish the best fit. Not all models are given the same weights in the fit, for the models both higher up the RGB ($(\textnormal{V}-\textnormal{V}_{\textnormal{\scriptsize{HB}}})$$<$--1) and at intermediate metallicities (--1.5$\leq$[Fe/H]$\leq$--0.5) we give higher weights since these models represent the best studied region of, and best matching models with, the empirical relation. Additionally, all models at the (extremely) low-metallicity regime we are particularly interested in (below [Fe/H]=$-2$) are also weighted more. On the other hand, the highest metallicity models (at [Fe/H]=$-0.5$) are weighted less, since these have higher uncertainties and fit less well the empirical relation. Several functional forms have been explored. The best fit was obtained by a simple linear relation in both luminosity and equivalent width, as in the classical empirical calibrations, with merely two extra terms. One term of equivalent width to the power -1.5 to account for the variations at low metallicities, one cross term to account for the slight non-linear slope, as given in Eq. \ref{newcalform}.

\begin{eqnarray}\label{newcalform}
\textnormal[\textnormal{Fe}/\textnormal{H}] = -2.87 + 0.195\times(\textnormal{V}-\textnormal{V}_{\textnormal{\scriptsize{HB}}}) + 0.458\times\textnormal{EW}_{(2+3)}  \nonumber \\
 - 0.913\times\textnormal{EW}_{(2+3)}^{-1.5} + 0.0155\times\textnormal{EW}_{(2+3)}\times(\textnormal{V}-\textnormal{V}_{\textnormal{\scriptsize{HB}}})
\end{eqnarray}  

\begin{table*}
\caption{The parameters of the observed stars used for verification of the modeling results at low metallicities}
\label{proplowmetcal}
\centering
\begin{tabular}{lllccccc}
\hline\hline
star & CaT reference & HR reference & T$_{\textnormal{\scriptsize{eff}}}$ (K) & log(g) (cm/s$^{2}$)& [Fe/H] HR (dex) & [Fe/H] CaT(new) (dex) \\
\hline
HD122563 & W. Aoki, priv. comm. & \citet{hond04} & 4570 & 1.1 & -2.77 & -2.66\\
HD110184 & J. Fulbright, priv. comm. & \citet{fulb00} & 4400 & 0.6 & -2.3 & -2.27\\
HD88609 & \citet{marr03} & \citet{fulb00} & 4450 & 0.6 & -2.9 & -2.97\\
HD4306 & J. Fulbright, priv. comm. & \citet{fulb00} & 4800 & 1.7 & -2.8 & -2.57\\
HD216143 & J. Fulbright, priv. comm. & \citet{fulb00} & 4525 & 1.0 & -2.1 & -2.10\\
CD -38 245 & VLT X-shooter commissioning & \citet{cayr04} & 4800 & 1.5 & -4.2 & -3.82\\
\hline
Boo I 1137 & \citet{norr08} & \citet{norr08} & - & - & -3.7 & -3.32 \\
\hline
\hline
\end{tabular}
\end{table*}

This relation is only calibrated for RGB stars above the HB and should thus not be applied to stars outside $-3 < (\textnormal{V}-\textnormal{V}_{\textnormal{\scriptsize{HB}}}) < 0$. We want to stress that this relation remains `empirical', in the sense that no theoretical arguments are used to find the best fitting formula. The corresponding fit is shown in Fig. \ref{newcalibfig_a}, it fits both the higher metallicity end (and in this regime the existing empirical relation) and the lower metallicity models well within their uncertainties. From Fig. \ref{newcalibfig_b}, which shows the residual of the fit in the right panel, it can be seen that the new calibration performs less well at the high-metallicity end of the calibration ([Fe/H]$\ga$--0.5), but even there the error is still within a typical observational error bar for [Fe/H]. We estimate the typical maximum error on the fitted parameters in Eq. \ref{newcalform} to be $\sim$8\%, on the basis of Monte-Carlo simulations of the uncertainties on (V-V$_{\textnormal{\scriptsize{HB}}}$) and the equivalent widths. These reasonably low error values convince us that the parameters in our new CaT calibration are quite robust to changes in our approximations.

Additionally, in Appendix \ref{absmag}, we describe the relation between [Fe/H], EW and M$_{\textnormal{\scriptsize{V}}}$ and M$_{\textnormal{\scriptsize{I}}}$, to enable the use of the CaT lines as a metallicity estimator for individual RGB stars in systems without a well-defined horizontal branch or for individual RGB field stars.

\subsection{Verifying the new calibration at low-metallicity}

\begin{figure}
\resizebox{\hsize}{!}{\includegraphics{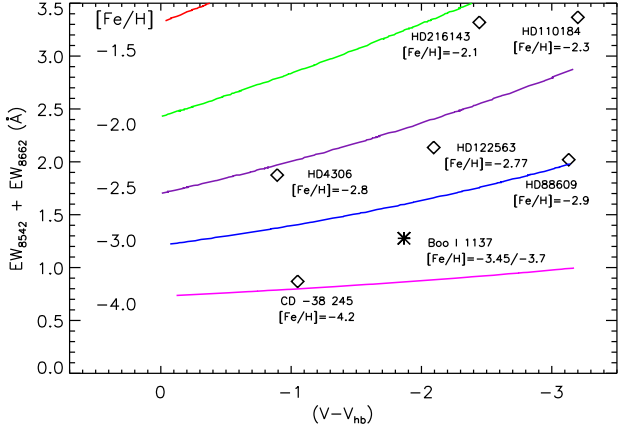}}
\caption{Plotted with our new CaT calibration (colored lines) are the well-studied RGB stars in the Milky Way halo (black diamonds) and one extremely low-metallicity star in Bo\"otes I (black asterisk) which are all described in Table \ref{proplowmetcal}. \label{plotlowmetstars}}
\end{figure} 

To verify the reliability of our models at very low-metallicities ([Fe/H]$\leq$--2.0), we have measured the CaT equivalent widths for six low-metallicity halo stars with existing high-resolution spectroscopic analyses. The properties of these stars are given in Table \ref{proplowmetcal}. Their CaT spectra are degraded to a resolution of R=6500, equal to the resolution of the synthetic spectra grid, and their CaT EWs are measured as described in Section \ref{emprel}. M$_{\textnormal{\scriptsize{V}}}$ for these halo stars is calculated from M$_{\textnormal{\scriptsize{bol}}}$ using the spectroscopically defined values for gravity and temperature and the bolometric correction of giant stars from \citet{alon99}. Additionally their height above the horizontal branch can be approximated using their spectroscopic metallicity and the relation from \citet{cate08}. In Fig. \ref{plotlowmetstars} the results of applying our new CaT calibration to these observations are shown. There is clearly a very good agreement. 

Additionally, one star from the Bo\"otes I dwarf galaxy as studied by \citet{norr08,norr09} using medium- and high-resolution spectroscopy is also plotted. For this star only one of the two strongest CaT lines could be measured by \citet{norr08} with confidence, and the total equivalent width for both lines was inferred using this single line and the observed ratio between the two lines from \citet{norr08,norr96}\footnote{From our synthetic modeling results we can check this value by evaluating the ratio in line strength between the two strongest CaT lines. These results are shown in Appendix \ref{EW23ratios}.} \citet{norr08} find [Fe/H]=--3.45 from medium-resolution spectroscopy using the Ca II H and K lines, which is very close to the value we deduce from the EW of the CaT line in the same spectrum, [Fe/H]=--3.32. In their subsequent high-resolution follow-up study [Fe/H] is measured directly from Fe lines, which gives [Fe/H]=--3.7 \citep{norr09}.

\subsection{The DART low-metallicity follow-up program}\label{followup}

\begin{figure*}
\centering
\includegraphics[width=17cm]{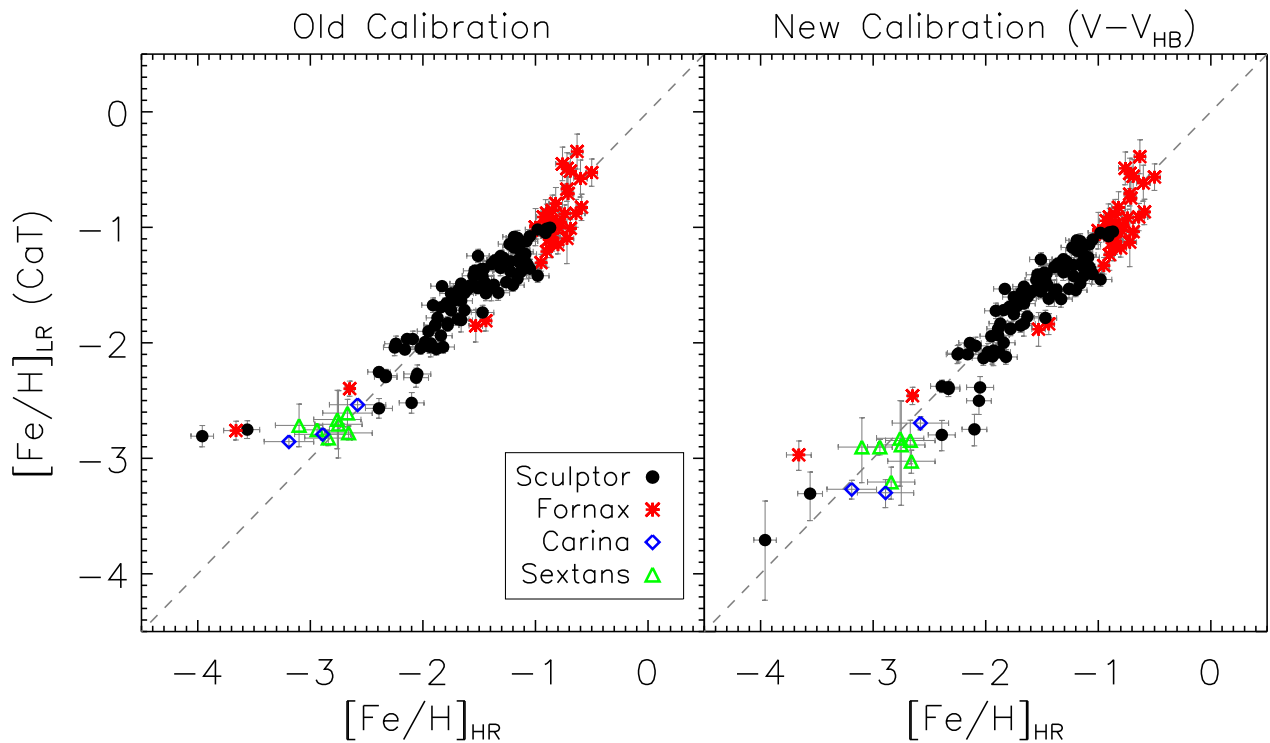}
\caption{A comparison between the CaT [Fe/H] and the [Fe/H] values as determined from high-resolution measurements for the same stars from the DART survey, for both the old (left panel) and new CaT calibrations (right panel). Stars from the galaxies Sculptor (black circles), Fornax (red asterisks), Carina (blue diamonds) and Sextans (green triangles) are included.\label{dartcalibration}}
\end{figure*}

As a complementary approach to determine if there are any extremely low metallicity stars in the dwarf galaxies, DART has undertaken a follow-up program to obtain HR spectroscopy using the Subaru Telescope High Dispersion Spectograph, the UVES spectrograph at VLT, and the MIKE spectrograph at Magellan for a sample of stars with CaT [Fe/H] < -2.5 \citep[][Tafelmeyer et al. in prep., Venn et al. in prep.]{aoki09}. These HR spectra were taken as an addition to the already existing HR spectra from the main program of DART using VLT/FLAMES with the GIRAFFE spectograph in Medusa mode \citep[][Letarte et al. in press, Hill et al. in prep., Venn et al. in prep.]{aoki09,tols09}. In these follow-up programmes several extremely low-metallicity stars have been found, with [Fe/H] values below --3.0 \citep[][Venn et al. in prep., Tafelmeyer et al. in prep.]{aoki09} and even around --4.0 (Tafelmeyer et al. in prep.). Figure \ref{dartcalibration} shows all the HR results, compared to their LR [Fe/H] values inferred from the CaT lines using both the old and the new calibration. In this figure the limiting range of the old (linear) calibration is also clearly visible, only the new calibration extends down to the lowest metallicities. For the lower metallicities the error bars become larger, due to the fact that the relations for different metallicities lie closer together and thus a similar error in equivalent width results in a larger error in [Fe/H]. Taking this into account, the new calibration appears to give an accurate prediction of the HR [Fe/H] values. 

There are however some stars showing a deviation larger than 1$\sigma$, where the most clear example is the extremely metal-poor star in Fornax. This deviation might be (partly) due to non-LTE effects or other deficiencies in the modeling of the HR spectrum in order to derive the stellar parameters and abundances for this really low-gravity star. Some support of this explanation is the fact that the agreement between the LR (CaT) and HR results improves when [\ion{Fe}{II}/H] is used instead of [\ion{Fe}{I}/H], most clearly for the extremely metal-poor Fornax star (Tafelmeyer et al. in prep.). Non-LTE effects are generally negligible for the dominant ionization state of \ion{Fe}{II} \citep[e.g.,][]{thev99,kraf03,mash09}, however for \ion{Fe}{I} the non-LTE effects are expected to be more significant in low-metallicity, low-gravity stars \citep[e.g.,][]{thev99,gehr01,mash09}.

\section{Alpha element dependence on the CaT lines}\label{alpha}
\begin{figure}
\resizebox{\hsize}{!}{\includegraphics{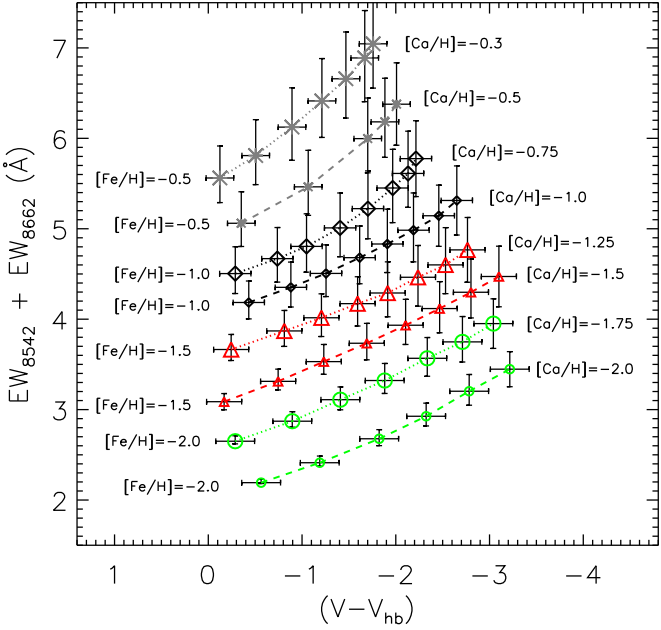}}
\caption{Synthetic spectra model predictions for [Ca/Fe]=+0.25 and [Ca/Fe]=0.0. Models with the same [Fe/H] are plotted using the same symbols and colors. However, the synthetic spectra models with [Ca/Fe]=+0.25 are plotted with larger symbols and are connected by dotted lines whereas the models with [Ca/Fe]=0.0 have smaller symbols and are connected by dashed lines. The [Fe/H] and [Ca/H] values for each set of models are indicated.\label{modemprel_alphas}}
\end{figure}

It is naturally expected that differences in [Ca/Fe] will significantly change the equivalent width of the CaT lines, and therefore alter the observed relation between equivalent width and [Fe/H]. The MARCS collaboration also provides model atmospheres where the $\alpha$ elements are not enhanced, but are kept similar to the ratio in the Sun for all models with [Fe/H] $\geq$ --2. Figure \ref{modemprel_alphas} shows that we recover this difference in our synthetic spectra. Qualitatively, the space between the enhanced and solar [Ca/Fe] models at equal [Fe/H] seems to agree with the step taken in abundance. The equivalent width in our grid of synthetic spectra does trace the Ca abundance. Since the strength of any line depends on the line opacity divided by the continuous opacity, it is expected that also other elements than Ca can affect the CaT line EW. If these elements contribute free electrons these can enhance the H$^{-}$ concentration and therefore affect the continuous opacity. Which element contributes most free electrons is dependent on the effective temperature and layer of the atmosphere of the star in consideration, but for cool stellar atmospheres in general the main sources of electrons are Mg, Fe, Si, Ca, Na, and Al \citep{shet09}. Since thus both Fe and some of the $\alpha$ elements have to be considered as important electron donors, this can significantly affect the dependence of the CaT EW on the [$\alpha$/Fe] ratio. Nevertheles we find that, within the abundance parameters we adopt in this study, the Ca abundance itself is by far the dominant factor determining the EW of the CaT line, as can be seen in Figure \ref{modemprel_alphas}. 

However, when comparing the sensitivity of the CaT equivalent width measurements of RGB stars in Sculptor and Fornax with high-resolution [Fe/H] and [Ca/H] measurements \citet{batt08b} find that the CaT equivalent width is actually a more robust estimator of Fe than Ca. This would suggest that it is therefore not advisable to use the CaT as a linear estimator for [Ca/H] \citep[see][and their Figs. 10 and 12]{batt08b}. This result is not expected from the theoretical expectations shown in Fig. \ref{modemprel_alphas}. There are several factors that may contribute to this apparent discrepancy between the modeling and observational results. First, in our grid of synthetic spectra we assume the relative abundance of Ca to the other $\alpha$ elements to stay constant, an assumption that not necessarily holds for all RGB stars. The extra free electrons donated by the other $\alpha$ elements might affect the strength of the CaT line through the continuous opacity as described above.  \citet{shet09} find that a 0.5 dex decrease in the electron contributors can increase the CaT line strength enough to mimic a Ca abundance increase of $\sim$0.2 or even $\sim$0.4 depending on the atmosphere parameters of the star. However, in Fornax we generally find that the important electron donors Fe and Mg are increased relatively to the values in our grid at similar [Ca/H], while we still measure the CaT lines to be too strong relative to the [Ca/H] measurements directly from \ion{Ca}{I} lines. This result clearly indicates that the effect of electron donors can not be driving the offset between [Ca/H] derived from CaT and HR \ion{Ca}{I} analyses. Second, the HR Ca abundances are usually derived from fewer \ion{Ca}{I} lines compared to the large number of HR \ion{Fe}{I} lines available and are therefore subject to larger observational errors. Third, the HR determination of the \ion{Ca}{I} abundance is subject to non-LTE effects \citep{mash07}. If non-LTE effects are included in the analysis of the HR spectra to determine the Ca abundances this might lead to a closer match in Ca abundances derived from the \ion{Ca}{I} and CaT lines. To investigate this more closely we have modeled both the \ion{Ca}{I} and CaT line strengths using abundances and atmospheric parameters from the well-studied halo star CD -38 245 \citep[e.g.,][]{cayr04} as a test case. The modeling, using the same models and techniques as described in Sect. \ref{non-LTE}, is performed for [Ca/Fe]=+0.4 in LTE and non-LTE to determine the offset from non-LTE to LTE abundances. The results are shown in Table \ref{Jonay_CaI}. It can be seen that the LTE approximation has an effect on the determination of the Ca abundance from the \ion{Ca}{I} lines (a difference of over 0.2 dex for the \ion{Ca}{I} line at $\lambda$ = 6162 $\AA$). In the LTE approximation, the agreement between the abundances derived from the \ion{Ca}{I} and CaT lines is very poor, CaT analysis results in a Ca abundance much greater than for \ion{Ca}{I}. In non-LTE, we are able to reproduce better all the measured Ca features in an extremely metal-poor star. The remaining discrepancy probably relates to the outer atmospheric layers that are not very well modeled - even in non-LTE. To fully resolve the discrepancy between the \ion{Ca}{I} and CaT results, one would have to properly explore the effects of specific details including line profile fitting and uncertainties in the stellar parameters. 

The stars in the dataset described by \citet{batt08b} which show the largest discrepancies in CaT compared to [Ca/H], are typically much more metal-rich RGB stars than CD -38 245. For this regime we have not modeled the non-LTE effects in \ion{Ca}{I} lines. Most of the discrepant stars are from the Fornax dwarf galaxy, where we were just able to target its brightest population of RGB stars due to the relatively large distance to this dwarf galaxy (see also Section \ref{lowmettails}). This means we are also statistically probing closer to the tip of the RGB, where we find the lower gravity stars for which the outer layers are more diffuse and thus more difficult to model - certainly assuming LTE.   

\begin{table}
\caption{Ca abundances for CD -38 245}
\label{Jonay_CaI}
\begin{tabular}{lccc}
\hline\hline
element & wavelength ($\AA$) & A(Ca) LTE & A(Ca) non-LTE \\
\hline
\ion{Ca}{I}& 4226.73 & 2.17 & 2.15 \\
\ion{Ca}{I}& 6162.18 & 2.12 & 2.35 \\
\ion{Ca}{II} & 8498.02 & 3.97 & 3.10 \\
\ion{Ca}{II} & 8542.09 & 3.42 & 2.83 \\
\ion{Ca}{II} & 8662.14 & 3.69 & 2.92 \\
\hline
\end{tabular}
\end{table}

\section{Implications for the DART survey}\label{implications}

Our new CaT-[Fe/H] calibration enables a more direct search for extremely metal-poor stars in existing datasets, like the large DART sample of CaT measurements in RGB stars of four classical dwarf galaxies (Sculptor, Fornax, Carina, and Sextans). These DART samples were observed in LR (R$\sim6500$) Medusa mode using the European Southern Observatory (ESO) VLT/FLAMES facility \citep{pasq02} and are described in \citet{helm06}, \citet{koch06} and \citet{batt06,batt08a,batt08b}.     

\begin{figure}
\includegraphics[width=\linewidth]{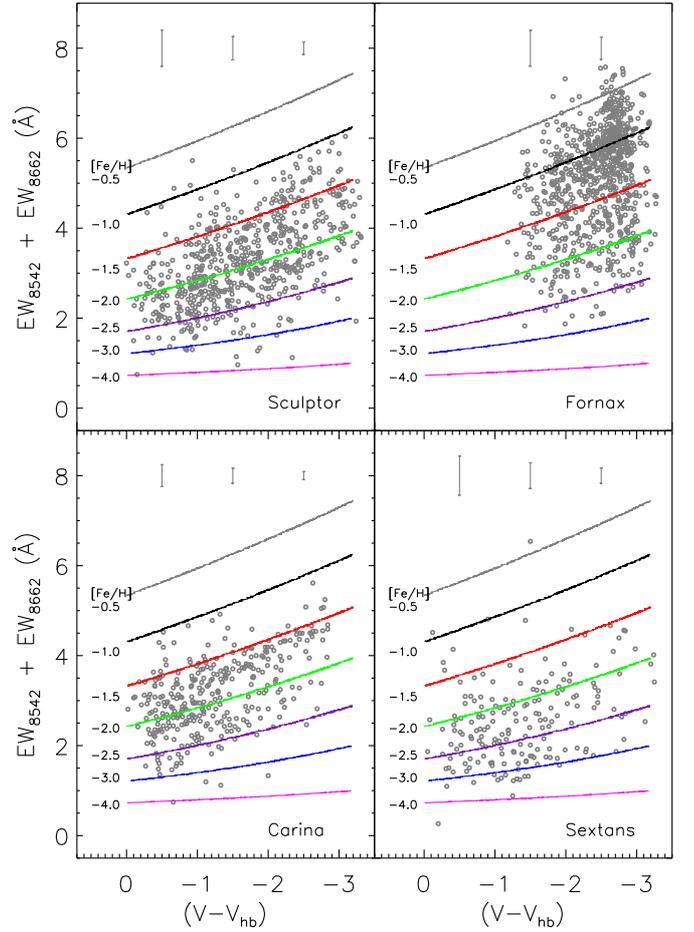}
\caption{Equivalent widths vs. the height above the HB for the DART dataset members of Sculptor, Fornax, Carina and Sextans (gray circles) in separate panels and the new CaT calibration at different [Fe/H] (colored lines). Typical error bars for the observed DART RGB stars for three luminosity bins are given in the upper part of each panel. \label{lowmetdartdata}}
\end{figure}

In Fig. \ref{lowmetdartdata} we overplot the complete low-resolution samples for the four galaxies observed in the DART program, Sculptor, Fornax, Carina and Sextans, along with the new CaT calibration (colored lines). All observed RGB stars, which are likely to be members, from DART are shown as small gray circles. All stars have a signal-to-noise ratio larger than 10 (per \AA), a velocity error smaller than 5 km/s and a velocity which is within 3$\sigma$ of the systemic radial velocity of the galaxy (for Fornax 2.5$\sigma$ is used, because of the larger contribution from Milky Way foreground). These criteria are identical to those applied by \citet{batt06,batt08a}. We use V$_{\textnormal{\scriptsize{HB}}}$ for the dwarf galaxies as given in \citet{irwi95}. For the errors in the sum of equivalent widths we use the results of \citet{batt08b} who found that the random error from repeated measurements of the low-resolution sample in the sum of the two broadest lines is well represented by $\sigma_{\textnormal{EW}_{\scriptsize{2+3}}}$$\approx6/$(S/N). Although this error can be quite extensive (mean error bars per galaxy and luminosity bin are shown in Fig. \ref{lowmetdartdata}), there are clearly a number of stars in these galaxies that are predicted to have [Fe/H]$<$--3. Our low-metallicity candidates need to be followed up with high-resolution spectroscopy to verify our prediction. A small sample of low-metallicity candidates has already been followed up with high-resolution spectroscopy within DART, these results are discussed in Sect. \ref{followup} and in \citet{aoki09}, Tafelmeyer et al. (in prep.) and Venn et al. (in prep.). The extremely metal-poor stars discovered within these DART follow-up programs in Sculptor, Fornax, Carina, and Sextans and the one extremely metal-poor Sculptor star discovered by \citet{kirb09} and \citet{freb09} already provide an independent confirmation of our predictions of the existence of extremely low-metallicity stars in these classical dwarf galaxies.

\subsection{Old and new calibration: A comparison}

\begin{figure}
\resizebox{\hsize}{!}{\includegraphics{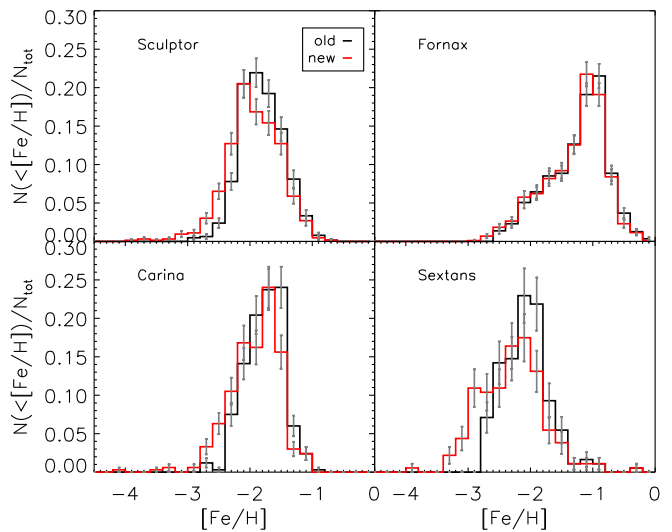}}
\caption{Distributions of [Fe/H] for the four dwarf spheroidal galaxies in the DART survey, using both the old calibration (Eq. \ref{emprel1} and \ref{emprel2}, black) and the new calibration (Eq. \ref{newcalform}, red) of the CaT lines. The error bars are Poissonian.\label{dwarfnewdistr}}
\end{figure}

It should be noted that the total number of stars affected by the new calibration in the DART CaT surveys is very small. In the current DART data set, just 2.5\% of the target stars have a metallicity below [Fe/H]=--2.5, using the calibration of \citet{batt08b}. Using the new calibration the fraction of low-metallicity stars increases somewhat to about 7.5\%. Still it is important to realize that $\sim$92.5\% of the stars in these systems are at a metallicity which is consistent with both the calibration of \citet{batt08b} and that given in Sect. \ref{newcalib} of this paper. This implies that the number of low-metallicity candidates remains very small compared to the total number of observed stars. This result is illustrated in Fig. \ref{dwarfnewdistr} where we compare the metallicity distributions for each galaxy using two calibrations of the CaT data: The calibration for dwarf galaxies published by \citet{batt08b} (in black), and the re-calibration presented here (in red). It is clear that the overall distributions of metallicities within these galaxies do not change significantly for the different CaT calibrations, the only clear differences are in the low-metallicity tails which have become more populated and more extended with the new CaT calibration.  

\subsection{The low-metallicity tails}\label{lowmettails}

\begin{figure}
\includegraphics[width=\linewidth]{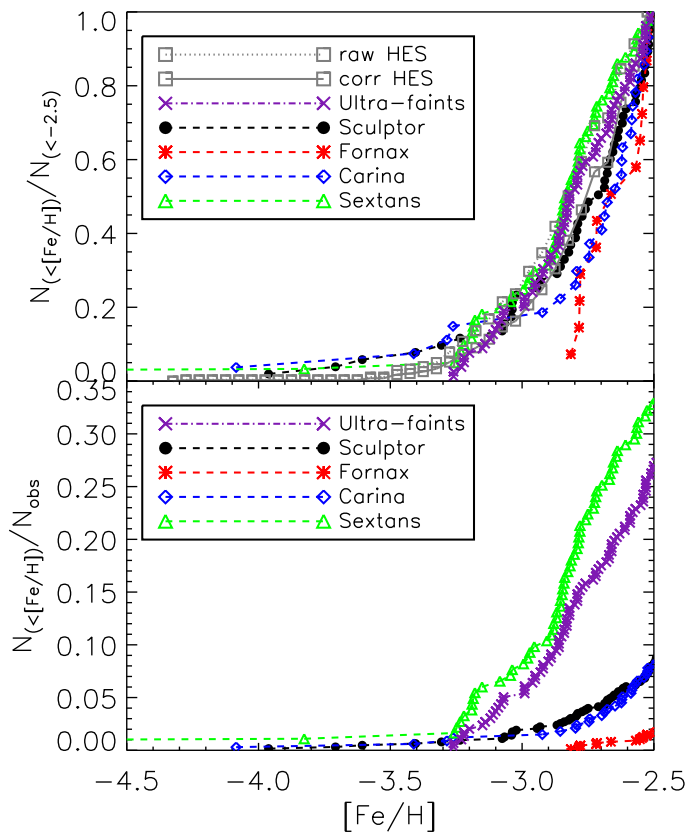}
\caption{The low-metallicity tails of the four classical dwarf galaxies studied by DART as obtained with the new calibration presented here. In the top panel these are normalized at [Fe/H]=--2.5 and compared to the low-metallicity tails from the HES survey (both corrected and raw) from \citet{scho09} and a compilation of ultra-faint dwarf galaxies \citep{kirb08}. The bottom panel displays the same low-metallicity tails for the classical- and ultra-faint dwarf galaxies, but normalized to the total number of observed stars in each system.\label{dwarfnewtail}}
\end{figure} 

Although their number is small relative to the total number of observed stars, the (extremely) metal-poor stars in each of these galaxies represents an important and interesting component of the overall stellar population. The low-metallicity population can reveal details of the chemical evolution of a particular galaxy and, by comparison, of the differences and similarities of the early evolutionary stages of galaxies. In Fig. \ref{dwarfnewtail} we show a comparison of the low-metallicity tails from individual stars, without application of binning or error estimates, in the four classical dwarf galaxies of DART using our new calibration for the CaT equivalent widths and in an ensemble of ultra-faint dwarf galaxies using low-resolution (R$\sim$6500) spectra over a large wavelength range compared with an extensive spectral library \citep{kirb08} and the HES survey of the Milky Way halo \citep{scho09} measuring [Fe/H] from a combination of indices for the \ion{Ca}{II} K and H$\alpha$ lines. Since the number of stars present, and observed, in these systems can vary quite a lot, a crucial step in comparing the low-metallicity tails is to normalize them at a certain metallicity or to a certain total number of measurements. In the upper panel of Fig. \ref{dwarfnewtail} all the low-metallicity tails have been normalized to 1 at [Fe/H] = --2.5. In this way, one can compare the shape of the metallicity tails below this value of [Fe/H] assuming all surveys are equally complete at this level. In the lower panel we have normalized to the total number of stars \textit{observed} in each system to give a feeling for the relative number of (extremely) low-metallicity stars observed. Ideally, one would like to be able to isolate the first generation of stars in all these galaxies and compare these populations. With our current understanding it is not possible to make such a clear distinction between samples, which is why we resort to the methods shown in Fig. \ref{dwarfnewtail}. However, with future larger samples of stars with accurate abundances in these galaxies one could envision making use of the position of the knee in the [$\alpha$/Fe] ratios for example \citep{tols09} to select the first generations of stars in each galaxy.

Before interpreting these metallicity tails further we stress that there are several caveats which prevent us from a detailed analysis of these results at face value. 

First, the errors on the individual measurements are large, especially at low [Fe/H] and/or for fainter stars (typical error bars are shown in Fig. \ref{lowmetdartdata}) and asymmetric. The asymmetry and dependence on metallicity arise from the fact that the relation between [Fe/H] and equivalent width is not linear at low metallicities, since the lines of equal metallicity are closer together. A symmetric error bar in the measured equivalent width, therefore results in a much larger error downwards than upwards in [Fe/H]. For instance, the one star in the Sextans dwarf galaxy that falls below the [Fe/H]=-4 calibration (see Fig. \ref{lowmetdartdata}) will be assigned a very low metallicity using the calibration, which causes the Sextans curve in Fig. \ref{dwarfnewtail} to stay above zero at [Fe/H]=--4.5. However, the upper 1$\sigma$ error in equivalent width allows a metallicity of [Fe/H]=--3.8. 

Second, there are several selection and/or sampling effects that are difficult to correct for. For example, the fact that not all stars have been observed in all of these systems. For instance, a much larger percentage of the Sextans stars have been observed than Fornax stars, although the total sample in Fornax is larger (it is a larger galaxy). Also, the absolute magnitude range of the populations targeted may not be comparable. From Fig. \ref{lowmetdartdata} it can clearly be seen that in Fornax, for instance, only stars well within 2 magnitudes from the tip of the RGB are observed due to its larger distance, while in the other galaxies the samples extend to fainter stars. Any luminosity bias that might remain in our new calibration would therefore result in a different bias for different galaxies in the extent and population of the low-metallicity tails. Even if one would observe the same range of magnitudes from the tip of the RGB for all galaxies, this could still introduce a bias, because of age and metallicity effects on the RGB for an extended star formation history. Additionally, some galaxies possess metallicity gradients \citep[e.g.,][]{tols04,batt06}, which means that the final metallicity distribution can also be greatly affected by the spatial sampling within a galaxy. 

Thus, even though from Fig. \ref{dwarfnewtail} one might be tempted to conclude that the low-metallicity tail of Fornax is different in both shape and number from the other classical dwarf galaxies and the halo and it must therefore have had a different chemical history in the earliest epochs, it is not at all clear whether such a conclusion is justified. It is very well possible that the low-metallicity tail of Fornax lies hidden under its very dominant, relatively young, metal-rich population and we have not been able to retrieve it because we sample a very small fraction of the total number of RGB stars in this galaxy. 

However, from the top panel of Fig. \ref{dwarfnewtail} it is clear that the significant difference between the metal-poor tail in the halo and in the dwarf galaxies from \citet{helm06} has disappeared. This also brings the metallicity distribution function for the classical dSphs in line with the results of \citet{kirb08} for the ultra-faint dwarf galaxies. The shapes of the low-metallicity tails, normalized at [Fe/H]=--2.5, are now much more similar.

\section{Conclusions}\label{conclusions}

It is important for our understanding of galactic chemical evolution to know whether there are just a few, and in this case how many, extremely low-metallicity ([Fe/H]$\leq -3$) stars in the classical dSph galaxies, or none at all. This can change our view of their early evolution and their subsequent role in galaxy formation and evolution. Subsequently, this will influence our view on the present-day satellite galaxies as possible templates for the building blocks which formed the Galactic halo and their relation to the more metal-poor ultra-faint dwarf galaxies.

Using a grid of synthetic spectra based on MARCS atmosphere models we have investigated the behavior of the three CaT absorption lines as metallicity indicators over a range of metallicities from [Fe/H]=--0.5 down to [Fe/H]=--4. Our models agree with the well-known observed linear relations at [Fe/H]$\geq$--2.0, although small deviations from linearity are found closer to the HB as were already predicted and observed \citep{pont04,carr07}. For [Fe/H]$<$--2.0, the relations of equal metallicity get closer together and are flatter as a function of absolute magnitude. This behavior was expected and is reflected in the change in the profile of the CaT lines, from wing-dominated to core-dominated as the metallicity drops. We have provided a new calibration taking this behavior into account, which we have successfully tested on several well-known low-metallicity giant stars. We have also investigated the role of the Ca abundance itself on these lines and found that although the lines ought to trace Ca instead of Fe, this often is not observed, probably due to inaccuracies in the determination of the HR Ca abundance. 

The new calibration of the CaT-[Fe/H] relation described here therefore provides a reliable updated method to search for low-metallicity stars using just the equivalent widths of two CaT lines. This method requires only a small wavelength range of the spectrum at low resolution and the absolute magnitude of the star. 

Applying the new calibration to the DART data sets shows that, although low in relative number, there are several extremely low-metallicity RGB candidates in these classical dwarf galaxies that deserve further study. This new calibration thus opens possibilities to study the (extremely) metal-poor tail that makes up only a small, but interesting, fraction of the dwarf galaxy stellar populations. 

\begin{acknowledgements}
We would like to warmly thank Jon Fulbright and Wako Aoki for sharing their CaT spectra of halo RGB stars with us and Bertrand Plez for making his linelists and Turbospectrum code available. We also thank Evan Kirby for supplying his data for the low-metallicity tail for the ultra-faint dwarfs and the referee, George Wallerstein, for useful suggestions that helped improve the paper. E.S., E.T., A.H., and T.d.B. gratefully acknowledge Netherlands Foundation for Scientific Research (NWO) and the Netherlands Research School for Astronomy (NOVA) for financial support. V.H. acknowledges the financial support of the Programme National Galaxies (PNG) of the Institut National des Sciences de l'Univers (INSU). A.H. was supported by the ERC-StG Galactica 240271. J.I.G.H. acknowledges financial support from EU contract MEXT-CT-2004-014265 (CIFIST) and from the Spanish Ministry project MICINN AYA2008-00695. This work has made use of BaSTI web tools.
\end{acknowledgements}

\begin{appendix}
\section{Absolute magnitude [Fe/H] and [Ca/H] calibrations}\label{absmag}

In numerous studies the equivalent widths of the CaT lines are studied in relation to the height of the stars above the HB. While this measure is very convenient for the study of stars in (galactic) globular clusters, one might also want to study systems which have no well-defined HB magnitude, or even no HB at all. In the main body of the paper we have chosen to define an extended calibration of the CaT equivalent width as a function of $(\textnormal{V}-\textnormal{V}_{\textnormal{\scriptsize{HB}}}$), which also enabled a direct comparison with the empirical relation of \citet{batt08b}. Here we also provide a calibration directly to (Johnson-Cousins) M$_{V}$ and M$_{I}$. Note that this also means that we can remove the assumption regarding the HB magnitude of our modeled `RGB stars'. An extra advantage of using  M$_{I}$ (instead of M$_{V}$) is that it is much less sensitive to age effects, and therefore provides more accurate metallicities for, in particular, younger populations of stars \citep{carr07}. We also provide calibrations for [Ca/H] instead of [Fe/H], using our full modeling grid of Table \ref{modelgrid} including the models with solar [$\alpha$/Fe] values. All calibrations use the same functional form given in Eq. \ref{newcalgenform}. The parameters for both [Fe/H] and [Ca/H] using either $(\textnormal{V}-\textnormal{V}_{\textnormal{\scriptsize{HB}}}$), M$_{\textnormal{\scriptsize{V}}}$, or M$_{\textnormal{\scriptsize{I}}}$ are given in Table \ref{parameters}. 

\begin{eqnarray}\label{newcalgenform}
\textnormal[\textnormal{Fe}/\textnormal{H}]\textnormal{\ or \ [\textnormal{Ca}/\textnormal{H}]} = \textnormal{a} + \textnormal{b}\times\textnormal{(Abs. mag.)} + \textnormal{c}\times\textnormal{EW}_{(2+3)} \nonumber \\
 + \textnormal{d}\times\textnormal{EW}_{(2+3)}^{-1.5} + \textnormal{e}\times\textnormal{EW}_{(2+3)}\times\textnormal{(Abs. mag.)}
\end{eqnarray}

\begin{table}
\caption{Best fitting parameters for the new calibrations}
\label{parameters}
\begin{tabular}{llr|llr}
\hline\hline
\multicolumn{3}{c|}{Calibrated for [Fe/H]} & \multicolumn{3}{c}{Calibrated for [Ca/H]}\\
\hline
Abs. mag. & parameter & value & Abs. mag. & parameter & value\\
\hline
(V-V$_{\textnormal{\scriptsize{HB}}}$) & a & -2.87 & (V-V$_{\textnormal{\scriptsize{HB}}}$) & a & -2.62 \\
 & b & 0.195 &  & b & 0.195\\
 & c & 0.458 &  & c & 0.457\\
 & d & -0.913 &  & d & -0.908\\
 & e & 0.0155 &  & e & 0.0146\\
\hline
M$_{\textnormal{\scriptsize{V}}}$ & a & -2.90 & M$_{\textnormal{\scriptsize{V}}}$ & a & -2.65\\
 & b & 0.187 &  & b & 0.185\\
 & c & 0.422 &  & c & 0.422\\
 & d & -0.882 &  & d & -0.876\\
 & e & 0.0133 &  & e & 0.0137\\
\hline
M$_{\textnormal{\scriptsize{I}}}$ & a & -2.78 & M$_{\textnormal{\scriptsize{I}}}$ & a & -2.53\\
 & b & 0.193 &  & b & 0.193\\
 & c & 0.442 &  & c & 0.439\\
 & d & -0.834 &  & d & -0.825\\
 & e & 0.0017 &  & e & 0.0013\\
\hline
\end{tabular}
\end{table}

\begin{figure}
\resizebox{\hsize}{!}{\includegraphics{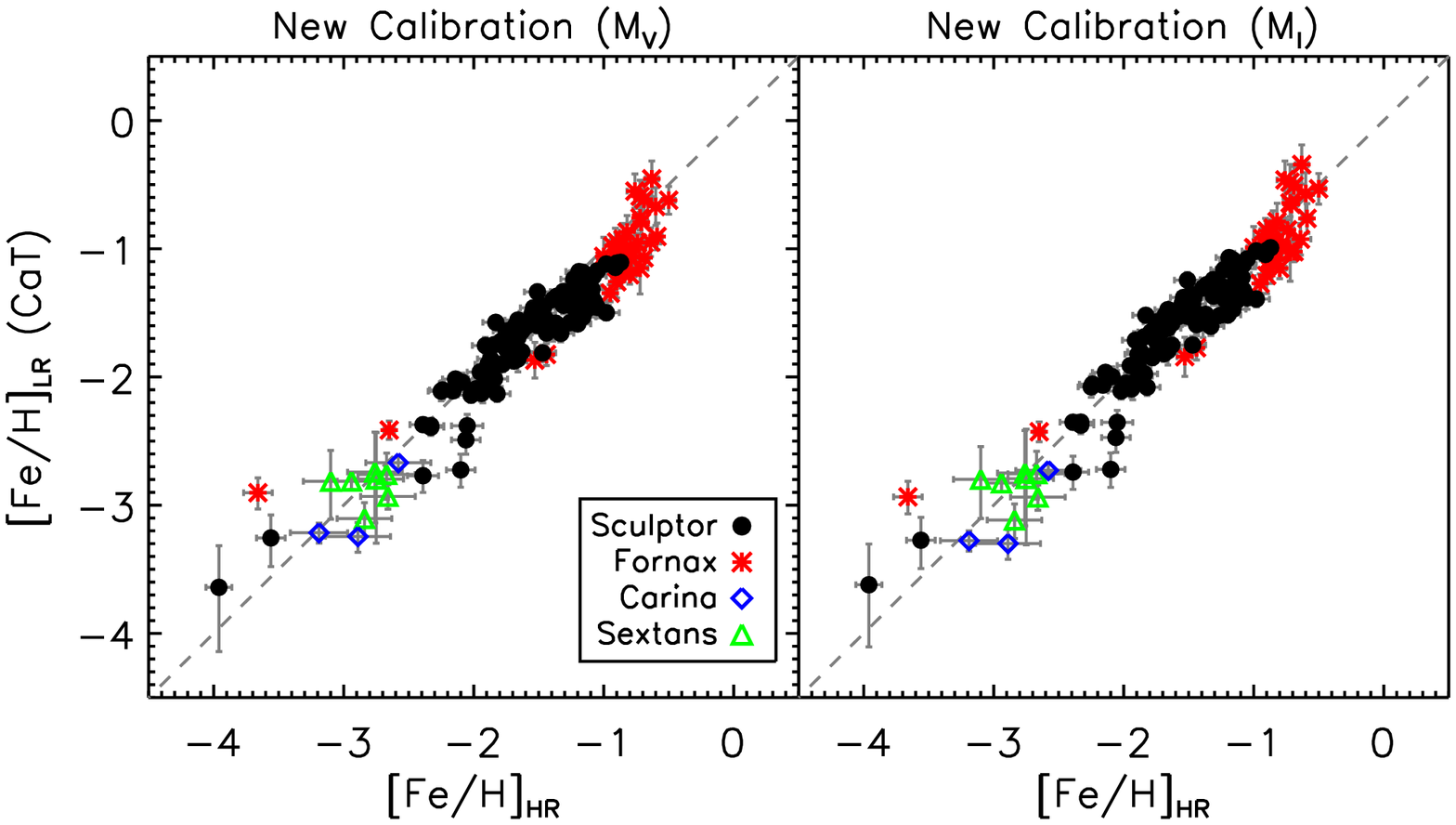}}\\
\resizebox{\hsize}{!}{\includegraphics{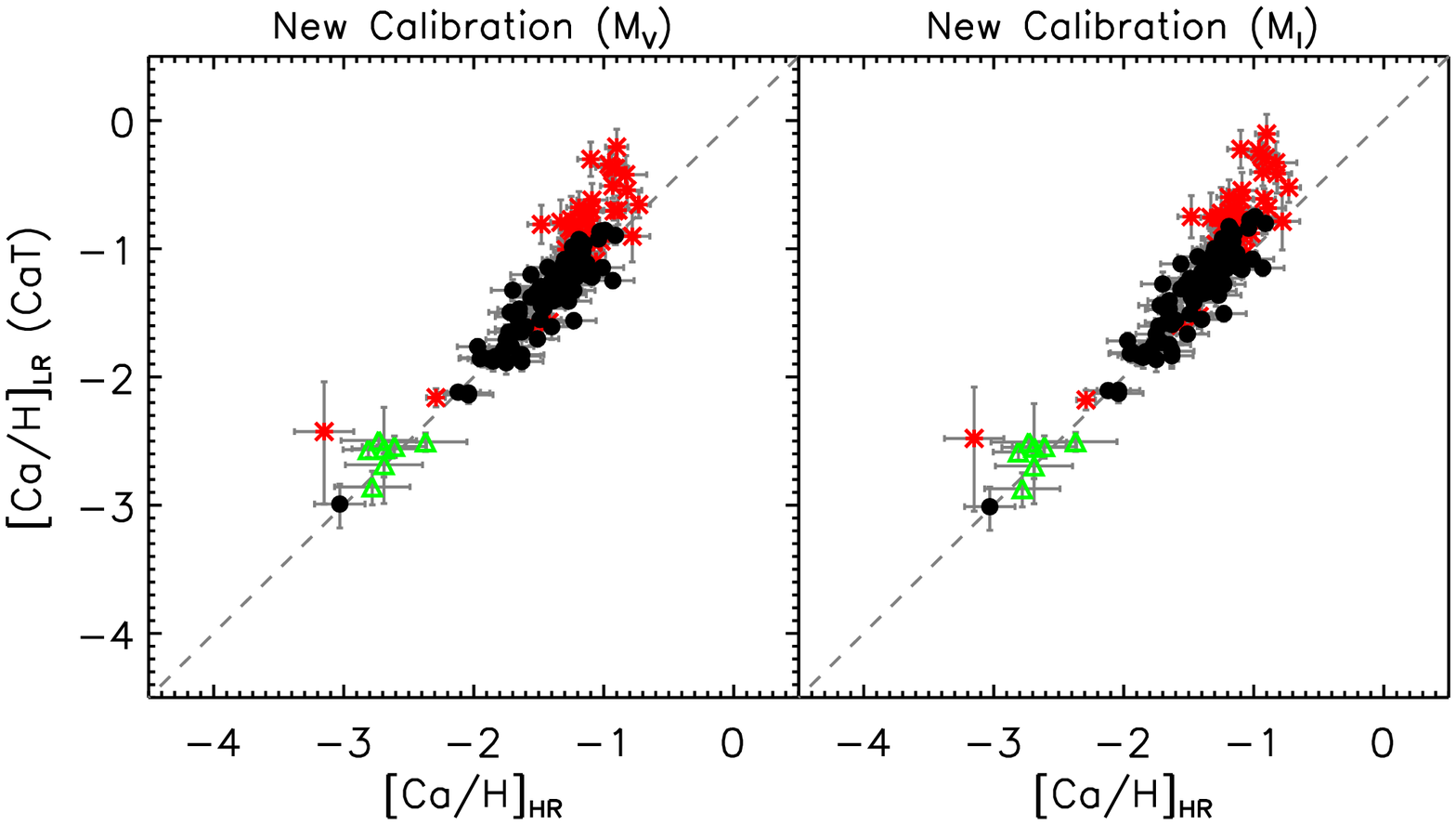}}
\caption{Top panels: Same as Fig. \ref{dartcalibration}, but now on the y-axis the new calibration for M$_{\textnormal{\scriptsize{V}}}$ and M$_{\textnormal{\scriptsize{I}}}$ are plotted. Bottom panels: The new calibration for [Ca/H] using either M$_{\textnormal{\scriptsize{V}}}$ or M$_{\textnormal{\scriptsize{I}}}$. \label{newcalibvvhbimag}}
\end{figure}

Since also these relations are only calibrated for RGB stars and above the HB, they should thus not be applied to stars outside $-3  < (\textnormal{V}-\textnormal{V}_{\textnormal{\scriptsize{HB}}}) <  0$, $-3 < M_{\textnormal{\scriptsize{V}}} < 0.8$, or $-4 < M_{\textnormal{\scriptsize{I}}} < 0$. In Fig. \ref{newcalibvvhbimag}, we show the results of the calibrations using M$_{\textnormal{\scriptsize{V}}}$ and M$_{\textnormal{\scriptsize{I}}}$ for the high-resolution DART sample for both [Fe/H] and [Ca/H]. Distance moduli for Sculptor, Fornax, Carina and Sextans are taken from \citet{kalu95}, \citet{rizz07}, \citet{mate98}, and \citet{mate95} respectively. The results do not change significantly depending on which absolute magnitude is used, as can also be seen from a comparison of the upper panels of Fig. \ref{newcalibvvhbimag} with Fig. \ref{dartcalibration}. It remains clear from Fig. \ref{newcalibvvhbimag}, that our new CaT calibrations based on the synthesized spectra trace the HR [Fe/H] values much better than HR [Ca/H], as was already discussed in Sect. \ref{alpha} of this paper and found by \citet{batt08b}. 

\section{Ratio of the 8542$\AA$ to the 8662$\AA$ CaT line}\label{EW23ratios}

\begin{figure}
  \resizebox{\hsize}{!}{\includegraphics{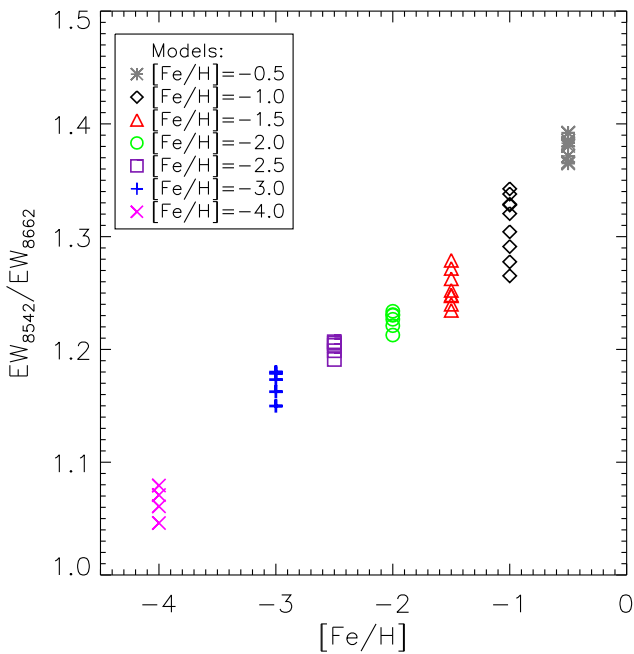}}
\caption{The ratio of the second (8542$\AA$) to the third (8662$\AA$) CaT line as a function of [Fe/H]. The color coding and symbols for the different [Fe/H] values are the same as in Fig. \ref{modemp_lowmet}.   \label{EW2to3ratio}}
\end{figure}

Another useful application of our models is the investigation of the ratios between the line strengths of the CaT lines, as a function of [Fe/H] and luminosity. This enables the determination of the summed equivalent width if just one of the two strongest lines are present or usable in the spectrum of a star. Additionally, investigation of the relative differences in development of the lines for various combinations of atmospheric parameters can give valuable insights into the line formation process. In Fig. \ref{EW2to3ratio} the ratios between the EWs of the second and third CaT lines, both of which are first corrected for non-LTE effects using Eqs. \ref{eqnlte1} and \ref{eqnlte2}, are plotted as a function of [Fe/H] for a synthetic `giant' spectrum. The variation in the ratio with absolute magnitude turns out to be very small as can be seen from the dispersion of the equal metallicity models. Although for the intermediate metallicity (--3$<$[Fe/H]$<$--1) spectra the ratio seems roughly constant with metallicity, the ratio changes significantly at higher ([Fe/H]=--0.5), or extremely low ([Fe/H]=--4) metallicities. We also find that the value for intermediate metallicities is somewhat lower than the value of 1.34 as measured by \citet{norr08}. Our mean ratio from the models is 1.27 if all models --2.5$\leq$[Fe/H]$\leq$--0.5 are taken into account (this particular range is chosen since it represents the metallicities of the vast majority of stars in the classical dwarf spheroidal galaxies). This value is confirmed in a comparison with the observed spectra from the DART dataset where a very similar mean value is found although the spread is much larger due to the observational uncertainties.

\end{appendix}

\bibliographystyle{aa}
\bibliography{bib13759}

\begin{thebibliography}{92}
\expandafter\ifx\csname natexlab\endcsname\relax\def\natexlab#1{#1}\fi

\bibitem[{{Alonso} {et~al.}(1999){Alonso}, {Arribas}, \&
  {Mart{\'{\i}}nez-Roger}}]{alon99}
{Alonso}, A., {Arribas}, S., \& {Mart{\'{\i}}nez-Roger}, C. 1999, \aaps, 140,
  261

\bibitem[{{Alvarez} \& {Plez}(1998)}]{alva98}
{Alvarez}, R. \& {Plez}, B. 1998, \aap, 330, 1109

\bibitem[{{Aoki} {et~al.}(2009){Aoki}, {Arimoto}, {Sadakane}, {Tolstoy},
  {Battaglia}, {Jablonka}, {Shetrone}, {Letarte}, {Irwin}, {Hill}, {Francois},
  {Venn}, {Primas}, {Helmi}, {Kaufer}, {Tafelmeyer}, {Szeifert}, \&
  {Babusiaux}}]{aoki09}
{Aoki}, W., {Arimoto}, N., {Sadakane}, K., {et~al.} 2009, \aap, 502, 569

\bibitem[{{Armandroff} \& {Da Costa}(1991)}]{arma91}
{Armandroff}, T.~E. \& {Da Costa}, G.~S. 1991, \aj, 101, 1329

\bibitem[{{Armandroff} \& {Zinn}(1988)}]{arma88}
{Armandroff}, T.~E. \& {Zinn}, R. 1988, \aj, 96, 92

\bibitem[{{Barklem} {et~al.}(2005){Barklem}, {Christlieb}, {Beers}, {Hill},
  {Bessell}, {Holmberg}, {Marsteller}, {Rossi}, {Zickgraf}, \&
  {Reimers}}]{bark05}
{Barklem}, P.~S., {Christlieb}, N., {Beers}, T.~C., {et~al.} 2005, \aap, 439,
  129

\bibitem[{{Battaglia} {et~al.}(2008a){Battaglia}, {Helmi}, {Tolstoy}, {Irwin},
  {Hill}, \& {Jablonka}}]{batt08a}
{Battaglia}, G., {Helmi}, A., {Tolstoy}, E., {et~al.} 2008a, \apjl, 681, L13

\bibitem[{{Battaglia} {et~al.}(2008b){Battaglia}, {Irwin}, {Tolstoy}, {Hill},
  {Helmi}, {Letarte}, \& {Jablonka}}]{batt08b}
{Battaglia}, G., {Irwin}, M., {Tolstoy}, E., {et~al.} 2008b, \mnras, 383, 183

\bibitem[{{Battaglia} {et~al.}(2006){Battaglia}, {Tolstoy}, {Helmi}, {Irwin},
  {Letarte}, {Jablonka}, {Hill}, {Venn}, {Shetrone}, {Arimoto}, {Primas},
  {Kaufer}, {Francois}, {Szeifert}, {Abel}, \& {Sadakane}}]{batt06}
{Battaglia}, G., {Tolstoy}, E., {Helmi}, A., {et~al.} 2006, \aap, 459, 423

\bibitem[{{Butler} \& {Giddings}(1985)}]{butl85}
{Butler}, K. \& {Giddings}, J.~R. 1985, Newsletter of Analysis of Astronomical
  Spectra 9

\bibitem[{{Carrera} {et~al.}(2007){Carrera}, {Gallart}, {Pancino}, \&
  {Zinn}}]{carr07}
{Carrera}, R., {Gallart}, C., {Pancino}, E., \& {Zinn}, R. 2007, \aj, 134, 1298

\bibitem[{{Castelli} \& {Kurucz}(2003)}]{cast03}
{Castelli}, F. \& {Kurucz}, R.~L. 2003, in IAU Symposium, Vol. 210, Modelling
  of Stellar Atmospheres, ed. {N.~Piskunov, W.~W.~Weiss, \& D.~F.~Gray}, 20P--+

\bibitem[{{Catelan} \& {Cort{\'e}s}(2008)}]{cate08}
{Catelan}, M. \& {Cort{\'e}s}, C. 2008, \apjl, 676, L135

\bibitem[{{Cayrel} {et~al.}(2004){Cayrel}, {Depagne}, {Spite}, {Hill}, {Spite},
  {Fran{\c c}ois}, {Plez}, {Beers}, {Primas}, {Andersen}, {Barbuy},
  {Bonifacio}, {Molaro}, \& {Nordstr{\"o}m}}]{cayr04}
{Cayrel}, R., {Depagne}, E., {Spite}, M., {et~al.} 2004, \aap, 416, 1117

\bibitem[{{Cenarro} {et~al.}(2001){Cenarro}, {Cardiel}, {Gorgas}, {Peletier},
  {Vazdekis}, \& {Prada}}]{cena01}
{Cenarro}, A.~J., {Cardiel}, N., {Gorgas}, J., {et~al.} 2001, \mnras, 326, 959

\bibitem[{{Cenarro} {et~al.}(2002){Cenarro}, {Gorgas}, {Cardiel}, {Vazdekis},
  \& {Peletier}}]{cena02}
{Cenarro}, A.~J., {Gorgas}, J., {Cardiel}, N., {Vazdekis}, A., \& {Peletier},
  R.~F. 2002, \mnras, 329, 863

\bibitem[{{Cohen}(1978)}]{cohe78}
{Cohen}, J.~G. 1978, \apj, 221, 788

\bibitem[{{Cohen} \& {Huang}(2009)}]{cohe09}
{Cohen}, J.~G. \& {Huang}, W. 2009, \apj, 701, 1053

\bibitem[{{Cole} {et~al.}(2004){Cole}, {Smecker-Hane}, {Tolstoy}, {Bosler}, \&
  {Gallagher}}]{cole04}
{Cole}, A.~A., {Smecker-Hane}, T.~A., {Tolstoy}, E., {Bosler}, T.~L., \&
  {Gallagher}, J.~S. 2004, \mnras, 347, 367

\bibitem[{{Da Costa} {et~al.}(2009){Da Costa}, {Held}, {Saviane}, \&
  {Gullieuszik}}]{daco09}
{Da Costa}, G.~S., {Held}, E.~V., {Saviane}, I., \& {Gullieuszik}, M. 2009,
  \apj, 705, 1481

\bibitem[{{Demarque} {et~al.}(2004){Demarque}, {Woo}, {Kim}, \& {Yi}}]{dema04}
{Demarque}, P., {Woo}, J., {Kim}, Y., \& {Yi}, S.~K. 2004, \apjs, 155, 667

\bibitem[{{Frebel} {et~al.}(2009){Frebel}, {Kirby}, \& {Simon}}]{freb09}
{Frebel}, A., {Kirby}, E., \& {Simon}, J.~D. 2009, ArXiv e-prints

\bibitem[{{Frebel} {et~al.}(2010){Frebel}, {Simon}, {Geha}, \&
  {Willman}}]{freb10}
{Frebel}, A., {Simon}, J.~D., {Geha}, M., \& {Willman}, B. 2010, \apj, 708, 560

\bibitem[{{Freeman} \& {Bland-Hawthorn}(2002)}]{free02}
{Freeman}, K. \& {Bland-Hawthorn}, J. 2002, \araa, 40, 487

\bibitem[{{Fulbright}(2000)}]{fulb00}
{Fulbright}, J.~P. 2000, \aj, 120, 1841

\bibitem[{{Fulbright} {et~al.}(2007){Fulbright}, {McWilliam}, \&
  {Rich}}]{fulb07}
{Fulbright}, J.~P., {McWilliam}, A., \& {Rich}, R.~M. 2007, \apj, 661, 1152

\bibitem[{{Fulbright} {et~al.}(2004){Fulbright}, {Rich}, \& {Castro}}]{fulb04}
{Fulbright}, J.~P., {Rich}, R.~M., \& {Castro}, S. 2004, \apj, 612, 447

\bibitem[{{Gehren} {et~al.}(2001){Gehren}, {Butler}, {Mashonkina}, {Reetz}, \&
  {Shi}}]{gehr01}
{Gehren}, T., {Butler}, K., {Mashonkina}, L., {Reetz}, J., \& {Shi}, J. 2001,
  \aap, 366, 981

\bibitem[{{Giddings}(1981)}]{gidd81}
{Giddings}, J.~R. 1981, PhD thesis, AA(Moscow state University)

\bibitem[{{Gustafsson} {et~al.}(2008){Gustafsson}, {Edvardsson}, {Eriksson},
  {J{\o}rgensen}, {Nordlund}, \& {Plez}}]{gust08}
{Gustafsson}, B., {Edvardsson}, B., {Eriksson}, K., {et~al.} 2008, \aap, 486,
  951

\bibitem[{{Helmi} {et~al.}(2006){Helmi}, {Irwin}, {Tolstoy}, {Battaglia},
  {Hill}, {Jablonka}, {Venn}, {Shetrone}, {Letarte}, {Arimoto}, {Abel},
  {Francois}, {Kaufer}, {Primas}, {Sadakane}, \& {Szeifert}}]{helm06}
{Helmi}, A., {Irwin}, M.~J., {Tolstoy}, E., {et~al.} 2006, \apjl, 651, L121

\bibitem[{{Hinkle} {et~al.}(2000){Hinkle}, {Wallace}, {Valenti}, \&
  {Harmer}}]{hink00}
{Hinkle}, K., {Wallace}, L., {Valenti}, J., \& {Harmer}, D. 2000, {Visible and
  Near Infrared Atlas of the Arcturus Spectrum 3727-9300 A} (ASP)

\bibitem[{{Honda} {et~al.}(2004){Honda}, {Aoki}, {Kajino}, {Ando}, {Beers},
  {Izumiura}, {Sadakane}, \& {Takada-Hidai}}]{hond04}
{Honda}, S., {Aoki}, W., {Kajino}, T., {et~al.} 2004, \apj, 607, 474

\bibitem[{{Irwin} \& {Hatzidimitriou}(1995)}]{irwi95}
{Irwin}, M. \& {Hatzidimitriou}, D. 1995, \mnras, 277, 1354

\bibitem[{{Ivans} {et~al.}(2001){Ivans}, {Kraft}, {Sneden}, {Smith}, {Rich}, \&
  {Shetrone}}]{ivan01}
{Ivans}, I.~I., {Kraft}, R.~P., {Sneden}, C., {et~al.} 2001, \aj, 122, 1438

\bibitem[{{Jones} {et~al.}(1984){Jones}, {Alloin}, \& {Jones}}]{jone84}
{Jones}, J.~E., {Alloin}, D.~M., \& {Jones}, B.~J.~T. 1984, \apj, 283, 457

\bibitem[{{Jorgensen} {et~al.}(1992){Jorgensen}, {Carlsson}, \&
  {Johnson}}]{jorg92}
{Jorgensen}, U.~G., {Carlsson}, M., \& {Johnson}, H.~R. 1992, \aap, 254, 258

\bibitem[{{Kaluzny} {et~al.}(1995){Kaluzny}, {Kubiak}, {Szymanski}, {Udalski},
  {Krzeminski}, \& {Mateo}}]{kalu95}
{Kaluzny}, J., {Kubiak}, M., {Szymanski}, M., {et~al.} 1995, \aaps, 112, 407

\bibitem[{{Kirby} {et~al.}(2009){Kirby}, {Guhathakurta}, {Bolte}, {Sneden}, \&
  {Geha}}]{kirb09}
{Kirby}, E.~N., {Guhathakurta}, P., {Bolte}, M., {Sneden}, C., \& {Geha}, M.~C.
  2009, \apj, 705, 328

\bibitem[{{Kirby} {et~al.}(2008){Kirby}, {Simon}, {Geha}, {Guhathakurta}, \&
  {Frebel}}]{kirb08}
{Kirby}, E.~N., {Simon}, J.~D., {Geha}, M., {Guhathakurta}, P., \& {Frebel}, A.
  2008, \apjl, 685, L43

\bibitem[{{Koch} {et~al.}(2008a){Koch}, {Grebel}, {Gilmore}, {Wyse}, {Kleyna},
  {Harbeck}, {Wilkinson}, \& {Wyn Evans}}]{koch08a}
{Koch}, A., {Grebel}, E.~K., {Gilmore}, G.~F., {et~al.} 2008a, \aj, 135, 1580

\bibitem[{{Koch} {et~al.}(2007b){Koch}, {Grebel}, {Kleyna}, {Wilkinson},
  {Harbeck}, {Gilmore}, {Wyse}, \& {Evans}}]{koch07b}
{Koch}, A., {Grebel}, E.~K., {Kleyna}, J.~T., {et~al.} 2007b, \aj, 133, 270

\bibitem[{{Koch} {et~al.}(2006){Koch}, {Grebel}, {Wyse}, {Kleyna}, {Wilkinson},
  {Harbeck}, {Gilmore}, \& {Evans}}]{koch06}
{Koch}, A., {Grebel}, E.~K., {Wyse}, R.~F.~G., {et~al.} 2006, \aj, 131, 895

\bibitem[{{Koch} {et~al.}(2008b){Koch}, {McWilliam}, {Grebel}, {Zucker}, \&
  {Belokurov}}]{koch08b}
{Koch}, A., {McWilliam}, A., {Grebel}, E.~K., {Zucker}, D.~B., \& {Belokurov},
  V. 2008b, \apjl, 688, L13

\bibitem[{{Koch} {et~al.}(2007a){Koch}, {Wilkinson}, {Kleyna}, {Gilmore},
  {Grebel}, {Mackey}, {Evans}, \& {Wyse}}]{koch07a}
{Koch}, A., {Wilkinson}, M.~I., {Kleyna}, J.~T., {et~al.} 2007a, \apj, 657, 241

\bibitem[{{Koch} {et~al.}(2009){Koch}, {Wilkinson}, {Kleyna}, {Irwin},
  {Zucker}, {Belokurov}, {Gilmore}, {Fellhauer}, \& {Evans}}]{koch09}
{Koch}, A., {Wilkinson}, M.~I., {Kleyna}, J.~T., {et~al.} 2009, \apj, 690, 453

\bibitem[{{Kraft} \& {Ivans}(2003)}]{kraf03}
{Kraft}, R.~P. \& {Ivans}, I.~I. 2003, \pasp, 115, 143

\bibitem[{{Krauss} \& {Chaboyer}(2003)}]{krau03}
{Krauss}, L.~M. \& {Chaboyer}, B. 2003, Science, 299, 65

\bibitem[{{Kupka} {et~al.}(2000){Kupka}, {Ryabchikova}, {Piskunov}, {Stempels},
  \& {Weiss}}]{kupk00}
{Kupka}, F.~G., {Ryabchikova}, T.~A., {Piskunov}, N.~E., {Stempels}, H.~C., \&
  {Weiss}, W.~W. 2000, Baltic Astronomy, 9, 590

\bibitem[{{Kurucz}(1993)}]{kuru93}
{Kurucz}, R. 1993, ATLAS9 Stellar Atmosphere Programs and 2 km/s grid.~Kurucz
  CD-ROM No.~13.~ Cambridge, Mass.: Smithsonian Astrophysical Observatory,
  1993., 13

\bibitem[{{Kurucz} {et~al.}(1984){Kurucz}, {Furenlid}, {Brault}, \&
  {Testerman}}]{kuru84}
{Kurucz}, R.~L., {Furenlid}, I., {Brault}, J., \& {Testerman}, L. 1984, {Solar
  flux atlas from 296 to 1300 nm} (National Solar Observatory Atlas, Sunspot,
  New Mexico: National Solar Observatory)

\bibitem[{{Leaman} {et~al.}(2009){Leaman}, {Cole}, {Venn}, {Tolstoy}, {Irwin},
  {Szeifert}, {Skillman}, \& {McConnachie}}]{leam09}
{Leaman}, R., {Cole}, A.~A., {Venn}, K.~A., {et~al.} 2009, \apj, 699, 1

\bibitem[{{Markwardt}(2009)}]{mark09}
{Markwardt}, C.~B. 2009, in Astronomical Society of the Pacific Conference
  Series, Vol. 411, Astronomical Society of the Pacific Conference Series, ed.
  {D.~A.~Bohlender, D.~Durand, \& P.~Dowler}, 251--+

\bibitem[{{Marrese} {et~al.}(2003){Marrese}, {Boschi}, \& {Munari}}]{marr03}
{Marrese}, P.~M., {Boschi}, F., \& {Munari}, U. 2003, \aap, 406, 995

\bibitem[{{Mashonkina} {et~al.}(2009){Mashonkina}, {Gehren}, {Shi}, {Korn}, \&
  {Grupp}}]{mash09}
{Mashonkina}, L., {Gehren}, T., {Shi}, J., {Korn}, A., \& {Grupp}, F. 2009,
  ArXiv e-prints

\bibitem[{{Mashonkina} {et~al.}(2007){Mashonkina}, {Korn}, \&
  {Przybilla}}]{mash07}
{Mashonkina}, L., {Korn}, A.~J., \& {Przybilla}, N. 2007, \aap, 461, 261

\bibitem[{{Mateo} {et~al.}(1995){Mateo}, {Fischer}, \& {Krzeminski}}]{mate95}
{Mateo}, M., {Fischer}, P., \& {Krzeminski}, W. 1995, \aj, 110, 2166

\bibitem[{{Mateo} {et~al.}(1998){Mateo}, {Hurley-Keller}, \& {Nemec}}]{mate98}
{Mateo}, M., {Hurley-Keller}, D., \& {Nemec}, J. 1998, \aj, 115, 1856

\bibitem[{{Mu{\~n}oz} {et~al.}(2006){Mu{\~n}oz}, {Majewski}, {Zaggia},
  {Kunkel}, {Frinchaboy}, {Nidever}, {Crnojevic}, {Patterson}, {Crane},
  {Johnston}, {Sohn}, {Bernstein}, \& {Shectman}}]{muno06}
{Mu{\~n}oz}, R.~R., {Majewski}, S.~R., {Zaggia}, S., {et~al.} 2006, \apj, 649,
  201

\bibitem[{{Norris} {et~al.}(1996){Norris}, {Freeman}, \& {Mighell}}]{norr96}
{Norris}, J.~E., {Freeman}, K.~C., \& {Mighell}, K.~J. 1996, \apj, 462, 241

\bibitem[{{Norris} {et~al.}(2008){Norris}, {Gilmore}, {Wyse}, {Wilkinson},
  {Belokurov}, {Evans}, \& {Zucker}}]{norr08}
{Norris}, J.~E., {Gilmore}, G., {Wyse}, R.~F.~G., {et~al.} 2008, \apjl, 689,
  L113

\bibitem[{{Norris} {et~al.}(2009){Norris}, {Yong}, {Gilmore}, \&
  {Wyse}}]{norr09}
{Norris}, J.~E., {Yong}, D., {Gilmore}, G., \& {Wyse}, R.~F.~G. 2009, ArXiv
  e-prints

\bibitem[{{Olszewski} {et~al.}(1991){Olszewski}, {Schommer}, {Suntzeff}, \&
  {Harris}}]{olsz91}
{Olszewski}, E.~W., {Schommer}, R.~A., {Suntzeff}, N.~B., \& {Harris}, H.~C.
  1991, \aj, 101, 515

\bibitem[{{Pasquini} {et~al.}(2002){Pasquini}, {Avila}, {Blecha}, {Cacciari},
  {Cayatte}, {Colless}, {Damiani}, {de Propris}, {Dekker}, {di Marcantonio},
  {Farrell}, {Gillingham}, {Guinouard}, {Hammer}, {Kaufer}, {Hill}, {Marteaud},
  {Modigliani}, {Mulas}, {North}, {Popovic}, {Rossetti}, {Royer}, {Santin},
  {Schmutzer}, {Simond}, {Vola}, {Waller}, \& {Zoccali}}]{pasq02}
{Pasquini}, L., {Avila}, G., {Blecha}, A., {et~al.} 2002, The Messenger, 110, 1

\bibitem[{{Pietrinferni} {et~al.}(2004){Pietrinferni}, {Cassisi}, {Salaris}, \&
  {Castelli}}]{piet04}
{Pietrinferni}, A., {Cassisi}, S., {Salaris}, M., \& {Castelli}, F. 2004, \apj,
  612, 168

\bibitem[{{Plez}(1998)}]{plez98}
{Plez}, B. 1998, \aap, 337, 495

\bibitem[{{Plez}(2008)}]{plez08}
{Plez}, B. 2008, Physica Scripta Volume T, 133, 014003

\bibitem[{{Pont} {et~al.}(2004){Pont}, {Zinn}, {Gallart}, {Hardy}, \&
  {Winnick}}]{pont04}
{Pont}, F., {Zinn}, R., {Gallart}, C., {Hardy}, E., \& {Winnick}, R. 2004, \aj,
  127, 840

\bibitem[{{Pritzl} {et~al.}(2005){Pritzl}, {Venn}, \& {Irwin}}]{prit05}
{Pritzl}, B.~J., {Venn}, K.~A., \& {Irwin}, M. 2005, \aj, 130, 2140

\bibitem[{{Rich} {et~al.}(2005){Rich}, {Corsi}, {Cacciari}, {Federici}, {Fusi
  Pecci}, {Djorgovski}, \& {Freedman}}]{rich05}
{Rich}, R.~M., {Corsi}, C.~E., {Cacciari}, C., {et~al.} 2005, \aj, 129, 2670

\bibitem[{{Rizzi} {et~al.}(2007){Rizzi}, {Held}, {Saviane}, {Tully}, \&
  {Gullieuszik}}]{rizz07}
{Rizzi}, L., {Held}, E.~V., {Saviane}, I., {Tully}, R.~B., \& {Gullieuszik}, M.
  2007, \mnras, 380, 1255

\bibitem[{{Rutledge} {et~al.}(1997a){Rutledge}, {Hesser}, \&
  {Stetson}}]{rutl97a}
{Rutledge}, G.~A., {Hesser}, J.~E., \& {Stetson}, P.~B. 1997a, \pasp, 109, 907

\bibitem[{{Rutledge} {et~al.}(1997b){Rutledge}, {Hesser}, {Stetson}, {Mateo},
  {Simard}, {Bolte}, {Friel}, \& {Copin}}]{rutl97b}
{Rutledge}, G.~A., {Hesser}, J.~E., {Stetson}, P.~B., {et~al.} 1997b, \pasp,
  109, 883

\bibitem[{{Sbordone} {et~al.}(2004){Sbordone}, {Bonifacio}, {Castelli}, \&
  {Kurucz}}]{sbor04}
{Sbordone}, L., {Bonifacio}, P., {Castelli}, F., \& {Kurucz}, R.~L. 2004,
  Memorie della Societa Astronomica Italiana Supplement, 5, 93

\bibitem[{{Sch{\"o}rck} {et~al.}(2009){Sch{\"o}rck}, {Christlieb}, {Cohen},
  {Beers}, {Shectman}, {Thompson}, {McWilliam}, {Bessell}, {Norris},
  {Mel{\'e}ndez}, {Ram{\'{\i}}rez}, {Haynes}, {Cass}, {Hartley}, {Russell},
  {Watson}, {Zickgraf}, {Behnke}, {Fechner}, {Fuhrmeister}, {Barklem},
  {Edvardsson}, {Frebel}, {Wisotzki}, \& {Reimers}}]{scho09}
{Sch{\"o}rck}, T., {Christlieb}, N., {Cohen}, J.~G., {et~al.} 2009, \aap, 507,
  817

\bibitem[{{Shetrone} {et~al.}(2003){Shetrone}, {Venn}, {Tolstoy}, {Primas},
  {Hill}, \& {Kaufer}}]{shet03}
{Shetrone}, M., {Venn}, K.~A., {Tolstoy}, E., {et~al.} 2003, \aj, 125, 684

\bibitem[{{Shetrone} {et~al.}(2001){Shetrone}, {C{\^o}t{\'e}}, \&
  {Sargent}}]{shet01}
{Shetrone}, M.~D., {C{\^o}t{\'e}}, P., \& {Sargent}, W.~L.~W. 2001, \apj, 548,
  592

\bibitem[{{Shetrone} {et~al.}(2009){Shetrone}, {Siegel}, {Cook}, \&
  {Bosler}}]{shet09}
{Shetrone}, M.~D., {Siegel}, M.~H., {Cook}, D.~O., \& {Bosler}, T. 2009, \aj,
  137, 62

\bibitem[{{Simon} \& {Geha}(2007)}]{simo07}
{Simon}, J.~D. \& {Geha}, M. 2007, \apj, 670, 313

\bibitem[{{Spinrad} \& {Taylor}(1969)}]{spin69}
{Spinrad}, H. \& {Taylor}, B.~J. 1969, \apj, 157, 1279

\bibitem[{{Spinrad} \& {Taylor}(1971)}]{spin71}
{Spinrad}, H. \& {Taylor}, B.~J. 1971, \apjs, 22, 445

\bibitem[{{Spite} {et~al.}(2005){Spite}, {Cayrel}, {Plez}, {Hill}, {Spite},
  {Depagne}, {Fran{\c c}ois}, {Bonifacio}, {Barbuy}, {Beers}, {Andersen},
  {Molaro}, {Nordstr{\"o}m}, \& {Primas}}]{spit05}
{Spite}, M., {Cayrel}, R., {Plez}, B., {et~al.} 2005, \aap, 430, 655

\bibitem[{{Suntzeff} {et~al.}(1993){Suntzeff}, {Mateo}, {Terndrup},
  {Olszewski}, {Geisler}, \& {Weller}}]{sunt93}
{Suntzeff}, N.~B., {Mateo}, M., {Terndrup}, D.~M., {et~al.} 1993, \apj, 418,
  208

\bibitem[{{Th{\'e}venin} \& {Idiart}(1999)}]{thev99}
{Th{\'e}venin}, F. \& {Idiart}, T.~P. 1999, \apj, 521, 753

\bibitem[{{Tolstoy} {et~al.}(2006){Tolstoy}, {Hill}, {Irwin}, {Helmi},
  {Battaglia}, {Letarte}, {Venn}, {Jablonka}, {Shetrone}, {Arimoto}, {Abel},
  {Primas}, {Kaufer}, {Szeifert}, {Francois}, \& {Sadakane}}]{tols06}
{Tolstoy}, E., {Hill}, V., {Irwin}, M., {et~al.} 2006, The Messenger, 123, 33

\bibitem[{{Tolstoy} {et~al.}(2009){Tolstoy}, {Hill}, \& {Tosi}}]{tols09}
{Tolstoy}, E., {Hill}, V., \& {Tosi}, M. 2009, \araa, 47, 371

\bibitem[{{Tolstoy} {et~al.}(2004){Tolstoy}, {Irwin}, {Helmi}, {Battaglia},
  {Jablonka}, {Hill}, {Venn}, {Shetrone}, {Letarte}, {Cole}, {Primas},
  {Francois}, {Arimoto}, {Sadakane}, {Kaufer}, {Szeifert}, \& {Abel}}]{tols04}
{Tolstoy}, E., {Irwin}, M.~J., {Helmi}, A., {et~al.} 2004, \apjl, 617, L119

\bibitem[{{Tolstoy} {et~al.}(2003){Tolstoy}, {Venn}, {Shetrone}, {Primas},
  {Hill}, {Kaufer}, \& {Szeifert}}]{tols03}
{Tolstoy}, E., {Venn}, K.~A., {Shetrone}, M., {et~al.} 2003, \aj, 125, 707

\bibitem[{{Venn} {et~al.}(2004){Venn}, {Irwin}, {Shetrone}, {Tout}, {Hill}, \&
  {Tolstoy}}]{venn04}
{Venn}, K.~A., {Irwin}, M., {Shetrone}, M.~D., {et~al.} 2004, \aj, 128, 1177

\bibitem[{{Walker} {et~al.}(2009b){Walker}, {Belokurov}, {Evans}, {Irwin},
  {Mateo}, {Olszewski}, \& {Gilmore}}]{walk09b}
{Walker}, M.~G., {Belokurov}, V., {Evans}, N.~W., {et~al.} 2009b, \apjl, 694,
  L144

\bibitem[{{Walker} {et~al.}(2009a){Walker}, {Mateo}, {Olszewski}, {Sen}, \&
  {Woodroofe}}]{walk09a}
{Walker}, M.~G., {Mateo}, M., {Olszewski}, E.~W., {Sen}, B., \& {Woodroofe}, M.
  2009a, \aj, 137, 3109

\bibitem[{{Yi} {et~al.}(2001){Yi}, {Demarque}, {Kim}, {Lee}, {Ree}, {Lejeune},
  \& {Barnes}}]{yi01}
{Yi}, S., {Demarque}, P., {Kim}, Y., {et~al.} 2001, \apjs, 136, 417

\end{thebibliography}

\end{document}